\documentclass[preprint,nofootinbib,aps,superscriptaddress,eqsecnum]{revtex4-1}
\pdfoutput=1
\usepackage{amsmath,amssymb,graphicx,subfigure}

\usepackage{float}
\usepackage{graphicx}
\usepackage{amsmath}
\usepackage{amsfonts}
\usepackage{amssymb}
\usepackage{hyperref}
\usepackage{subfigure}
\usepackage{slashed}
\usepackage{setspace} 
\usepackage{caption,ulem}
\usepackage{color}
\allowdisplaybreaks
\def\bea{\begin{eqnarray}}
	\def\eea{\end{eqnarray}}
\def\be{\begin{equation}}
	\def\ee{\end{equation}}
\def\nn{\nonumber}

\newcommand{\beq}{\begin{equation}}
	\newcommand{\eeq}{\end{equation}}

\def\nn{\nonumber}

\usepackage[utf8]{inputenc}

\DeclareUnicodeCharacter{2212}{-}

\def\nn{\nonumber}

\begin{document}
	\title{Five-zero texture in neutrino-dark matter model within the framework of minimal extended seesaw}
	
	\author{Pritam Das}
	\email[Email Address: ]{prtmdas9@gmail.com}
	\affiliation{Department of Physics, Tezpur University, Assam, India-784028}  
		\affiliation{Department of Physics, Indian Institute of Technology, Assam, India-781039}  
	\author{Mrinal Kumar Das}
	\email[Email Address: ]{mkdas@tezu.ernet.in}
	\affiliation{Department of Physics, Tezpur University, Assam, India-784028} 
	
	\author{Najimuddin Khan}
	\email[Email Address: ]{khanphysics.123@gmail.com}
	\affiliation{School of Physical Sciences, Indian Association for the Cultivation of Science
		2A $\&$ 2B, Raja S.C. Mullick Road, Kolkata 700032, India}
\affiliation{School of physics, Institute for Research in Fundamental Sciences (IPM), P.O.Box 19395-5531, Tehran, Iran.	}

	\begin{abstract}
		We study a model of neutrino and dark matter within the framework of a minimal extended seesaw (MES). This model is based on $A_4$ flavour symmetry along with the discrete $Z_3\times Z_4$ symmetry to stabilize the dark matter and construct desired mass matrices for neutrino mass. Five-zero textures are imposed in the final $4\times4$ active-sterile mass matrix, which significantly reduces the free parameter in the model. Three right-handed neutrinos were considered, two of them have degenerate masses which help us to achieve baryogenesis via resonant leptogenesis. A singlet fermion (sterile neutrino) with mass $\sim\mathcal{O}$(eV) is also considered, and we are able to put bounds on active-sterile mixing parameters via neutrino oscillation data. Resonant enhancement of lepton asymmetry is studied at the TeV scale, where we discuss a few aspects of baryogenesis considering the flavour effects. The possibility of improvement in effective mass from $0\nu\beta\beta$ in the presence of a single generation of sterile neutrino flavour is also studied within the fermion sector. In the scalar sector, the imaginary component of the complex singlet scalar ($\chi$) is behaving as a potential dark matter candidate and simultaneously the real part of the complex scalar is associated with the fermion sector for sterile mass generation.
\end{abstract}

\maketitle
\section{Introduction}
In the journey of theoretical and experimental signs of progress, currently, we are at a golden period in particle physics. The discovery of Higgs boson \cite{Chatrchyan:2012xdj,Aad:2012tfa}, neutrino mass \cite{An:2012eh,Abe:2011fz,Abe:2016nxk}, the existence of dark matter \cite{Jungman:1995df,Bertone:2004pz} have glorified the field in the last two decades. Several experimental signs of progress on neutrino are running which are expected to bring significant contributions in the field of new physics, and recent light neutrino results are shown in table \ref{tab:d1}. Along with the three generations of active neutrinos, eV scale sterile neutrinos { catch} serious attention over time. LSND reported for the first time about the anomalies that they observed during the appearance of $\bar{\nu_{\mu}}\rightarrow\bar{\nu_e}$ channel \cite{Athanassopoulos:1996ds,Aguilar:2001ty}. Later MiniBooNE has also reported the presence of sterile neutrinos at 6.0$\sigma$ CL \cite{Aguilar-Arevalo:2018gpe}. At the same time, the Gallium experiments GALLEX and SAGE \cite{Hampel:1998xg, Abdurashitov:1999zd} and several observations of reactor antineutrino fluxes at short baseline experiment \cite{Collin:2016rao,Mention:2011rk,Mueller:2011nm} also detected such kind of anomalies with eV scaled mass splitting. 
{However, at the current time exact generations or mass scales of sterile neutrinos are still unclear and lots of studies were done so far with different sterile neutrino masses under various assumptions \cite{Dodelson:1993je,Ruchayskiy:2012si,Abazajian:2012ys,Abazajian:2017tcc, Adhikari:2016bei}}. Sterile neutrinos with the broad mass spectrum are highlighted in many BSM frameworks naturally, as long as there is minimal mixing with the active neutrinos \cite{Atre:2009rg,Deppisch:2015qwa, Adhikari:2016bei}.
The cosmological effect of sterile neutrinos can be observed if there were a sufficient amount of sterile neutrinos produced via the mixing with the active neutrinos in the early Universe. Currently, not only oscillation experiments are looking for eV scale sterile neutrinos, but the tritium decay experiment KATRIN has also upgraded itself to be sensitive towards the different generations of neutrinos \cite{Aker:2019uuj}. Moreover, observational cosmology also puts reasonable limits on the number and mass of relativistic neutrino states in the early Universe \cite{Hinshaw:2012aka, Ade:2015xua}. Hence, both the cosmological and laboratory-based studies find tremendous success in the behavioural study of sterile neutrinos \cite{Giunti:2019aiy,Boser:2019rta}. 

Problems with neutrino mass generation, baryon asymmetry of the Universe, absolute neutrino mass along with dark matter studies need an obvious extension beyond the standard model (BSM) frameworks. The first two puzzles are perhaps resolved by introducing right-handed (RH) neutrinos in the scenario. There are various seesaw mechanisms to address active-sterile mass generation simultaneously \cite{Laine:2008pg, Abada:2014zra}, however, the minimally extended version of the type-I seesaw has gained attention over time. This framework is popularly known as minimal extended seesaw (MES) \cite{Barry:2011wb,Zhang:2011vh,Nath:2016mts,Das:2018qyt,BhupalDev:2012jvh}. This framework is very convenient to address the mass generation for active as well as sterile neutrinos, and it covers a wide spectrum of sterile mass (eV to keV) \cite{Das:2019kmn,Benso:2019jog}. We exactly follow this framework in the present paper and study active-sterile mixing parameters using texture zero methods.

The analysis of neutrino phenomenology under the seesaw mechanism involves many more parameters than can be measured from the neutrino masses and mixing; hence, the situation becomes challenging to study. Texture zero in the final neutrino mass matrix implies some of the elements are much smaller than the other elements or zero, eventually, the number of free parameters is significantly reduced \cite{Kageyama:2002zw, Zhang:2013mb,Borah:2016xkc}. Flavour symmetries are widely used in neutrino phenomenology and play a crucial role in texture realization by reducing free parameters in the neutrino mass matrix \cite{Borah:2015vra,Babu:2009fd,Borah:2016xkc, Kageyama:2002zw, Dziewit:2016qri}.
Flavon fields introduced under flavour symmetries to describe neutrino mixing phenomenology are customarily SM singlets. Under specific flavour symmetry groups, the transformation behaviours of the flavons may vary. We have considered discrete $A_4$ flavour symmetry, under which flavons acquire specific vacuum alignments after symmetry breaking along a particular direction. 
In the (3+1) situation, there are ten elements in the $4\times4$ symmetric mass matrix, and 16 parameters (4 masses, six mixing angles and 6 phases). Typically, texture zero studies connect these parameters through zeros, and there we get definite bounds on them. Recent studies \cite{Borah:2017azf,Sarma:2018bgf,Zhang:2013mb,Borah:2016xkc} on this $4\times4$ mass have predicted a various number of zeros with better accuracy on different aspects. At present, we proposed to explore the possibility with a maximum of five zeros in the final mass matrix and study active-sterile mixing phenomenology. 

The baryon asymmetry of the current Universe is well addressed in the literature under a different mechanism. Thermal leptogenesis in trivial type-I seesaw is quite popular, due to the presence of heavy right-handed (RH) neutrinos \cite{Davidson:2008bu,Buchmuller:2004tu,Frossard:2012pc,Joshipura:2001ya}. The lightest RH neutrino decays out of equilibrium, satisfying Sakharov conditions \cite{Sakharov:1967dj} and simultaneously giving mass to the light neutrinos. However, thermal leptogenesis restricts the lower bound on the decaying RH neutrino ($M_{N_1}\ge10^9$ GeV). Such higher masses are not discoverable for recent/ongoing experiments, thus studies of low-scaled leptogenesis are preferable in the current situation. Resonant leptogenesis (RL) is { popular in low scale} \cite{Pilaftsis:2003gt, Hambye:2001eu, Cirigliano:2006nu,Chun:2007vh,Kitabayashi:2007bs,Bambhaniya:2016rbb}, which allow a nearly degenerate mass spectrum of RH neutrinos at TeV scale. In RL, the leptonic asymmetry gets resonantly enhanced up to the order of unity by the Majorana neutrino self-energy effect \cite{Liu:1993tg}, when the mass splitting between the RH neutrinos is of the order of the decay rates ($\Delta M\sim \Gamma$). As a result, in the thermal RL, the extra Majorana neutrino mass scale can be considered as low as the electroweak scale \cite{Pilaftsis:2003gt} while satisfying agreement with the neutrino oscillation data. 

Baryon asymmetry is generated when the RH neutrino decays out of equilibrium, and the abundance of RH neutrino along with the lepton generations are determined by the Boltzmann equation. There is the quantum approach of the Boltzmann equation, which is based on the Schwinger-Keldysh Closed Time Path (CTP) formulation \cite{Iso:2013lba}. Within this formulation, one needs to derive quantum field-theoretic analogues of the Boltzmann equations, known as Kadanoff-Baym (KB) equations \cite{kadanoff2018quantum, Garny:2011hg, BhupalDev:2014oar,Kartavtsev:2015vto} describing the non-equilibrium time-evolution of the two-point correlation functions. These time integrals of KB equations ensure non-Markovian which allows studying the history of the system as a memory effect. 
These equations are consistent for all flavour and thermal effects. In part works, a non-equilibrium perturbative
thermal field theory \cite{Millington:2012pf} was developed, which defines physically meaningful particle number densities from the Noether charge. { We followed the semi-classical approaches by \cite{Dev:2014laa, DeSimone:2007edo, Pilaftsis:2003gt, Deppisch:2010fr} to work out the flavor effect in the RL scenario.}

From the neutrinoless double beta decay ($0\nu\beta\beta$) study, if the Majorana nature of the neutrino can be substantiated, one can give a conclusive remark on absolute neutrino mass.
Lepton number is extensively violated in the $0\nu\beta\beta$ processes by creating a pair of electrons. Established results of the lepton number violation (LNV) process supported by the existing theoretical scenario and the $0\nu\beta\beta$ study allow leptons to take part in the process of matter-antimatter asymmetry of our Universe. Hence, the observation of this process is essential for demonstrating the baryogenesis idea \cite{Cline:2006ts} via lepton number violation. A rich literature  with SM neutrinos on $(0\nu\beta\beta)$ are reported in \cite{Bilenky:2004wn,Bilenky:2012qi,Agostini:2018tnm,Borgohain:2017akh,Awasthi:2013ff}. It is now clear that the addition of sterile neutrino with the SM particles can lead us to a broad range of new physics phenomenology \cite{Barry:2011wb,Abada:2018qok, Das:2019kmn}. Thus, motivated by these studies, we focus on adding an eV scaled sterile neutrino in the LNV process and study their consequences in this $(3+1)$ 5-zero texture work.  

The origin of dark matter and neutrino mass in the current situation is a puzzle, and there are various BSM frameworks at the tree level and loop level to address them under a single roof \cite{Hambye:2009pw, Magana:2012ph, Khan:2017xyh, Das:2014fea, Das:2019ntw}. We consider a minimal model with a complex scalar singlet $\chi=(\chi^R+i\chi^I)$, which takes part in fermion as well as scalar sector. A $Z_4$ symmetry is chosen in such a fashion that the complex scalar singlet remains odd, which restricts its mixing with other SM particles. In the fermion sector, { it couples} to the additional singlet neutral fermion; sterile neutrino and RH neutrino, which help to generate masses of the sterile and active neutrinos and other experimental neutrino variables. On the other hand, in the scalar sector, the lightest component of $\chi$ (the imaginary part, $\chi^I$) serves as a viable WIMP (weakly interacting massive particle) dark matter candidate. Other scalar singlets and triplets in the model are decoupled from the dark matter analysis due to the large mass; nonetheless, they take active participation in neutrino mass generation and the baryogenesis process. Our main aim is to keep all these particles in the chosen model to explain neutrino parameters using the texture zero methods, dark matter as well as baryogenesis and neutrinoless double beta decay altogether. It was not done previously in the literature, which motivates us to carry out this detailed study.

We organized this paper as follows: In section \ref{sec2}, we discussed the complete theoretical framework and the model description in subsequent subsections. In the sub-sections of section \ref{sec3}, we discuss various constraints and numerical approaches for minimal extended seesaw and various parametrization, baryogenesis, neutrinoless double beta decay and dark matter respectively. In section \ref{sec4}, we discuss the numerical results of our work separately in each sub-sections, and finally, we concluded our work in section \ref{sec5}.
\begin{table}[t]
	\begin{tabular*}{\columnwidth}{@{\extracolsep{\fill}}|l|l|l|@{}}
		\hline
		Parameters&bfp$\pm 1\sigma$&Normal Ordering\\
		\hline
		$\Delta m^{2}_{21}[10^{-5}eV^2]$&$7.42^{+0.21}_{-0.20}$&6.82-8.04\\
		$\Delta m^{2}_{31}[10^{-3}eV^2]$&$2.517^{+0.026}_{-0.028}$&2.435-2.598\\
		$sin^{2}\theta_{12}/10^{-1}$&$3.04^{+0.12}_{-0.12}$&2.69-3.43\\
		$sin^{2}\theta_{13}/10^{-2}$&$2.219^{+0.062}_{-0.063}$&2.032-2.41\\ 
		$sin^{2}\theta_{23}/10^{-1}$&$5.73^{+0.16}_{-0.20}$&4.15-6.16\\
		$\delta_{13}/\circ$&$197^{+27}_{-24}$&120-369\\ \hline
	\end{tabular*}
	\caption{Recent  NuFIT 5.0 (2020) results for active neutrino oscillation parameters with best-fit and the latest global fit $3\sigma$ range for normal ordering mode \cite{Esteban:2020cvm}.}\label{tab:d1}
\end{table}
\section{Theoretical framework}\label{sec2}
\subsection{Choice of 5-zero texture}\label{s21}
Recent studies \cite{Borah:2016xkc,Borah:2017azf} show a maximum of five zeros in the $4\times4$ active-sterile mass matrix were possible and beyond five zeros ($i.e.,$ six zeros in the $4\times4$ mass matrix), it fails to hold the latest bounds on the mixing parameters. In this work also, we focus on the maximum possible zero in the final $4\times4$ active-sterile mass matrix and study phenomenological consequences. In the five zeros texture, there are $^{10}C_5=252$ possibilities of zeros and 246 among them are ruled out due to the condition $(M_{\nu})_{i4}\ne0$ (with $i=e,\mu,\tau,4$)\footnote{Earlier texture zero studies in the 3 + 1 neutrino framework \cite{Ghosh:2013nya}, has shown that, the zeros in the active and extended sterile	sector at the same time is phenomenologically disallowed.}.  Hence, we are left with six choices of mass matrices with zeros in the active sector. They are as follows,
\begin{eqnarray}\label{txz}
\nn
T_1=\begin{pmatrix}
X&0&0&X\\
0&0&0&X\\
0&0&0&X\\
X&X&X&X\\
\end{pmatrix};~\quad
T_2=\begin{pmatrix}
0&0&0&X\\
0&X&0&X\\
0&0&0&X\\
X&X&X&X\\
\end{pmatrix}; ~\quad
T_3=\begin{pmatrix}
0&X&0&X\\
X&0&0&X\\
0&0&0&X\\
X&X&X&X\\
\end{pmatrix};\\
T_4=\begin{pmatrix}
0&0&X&X\\
0&0&0&X\\
X&0&0&X\\
X&X&X&X\\
\end{pmatrix};~\quad
T_5=\begin{pmatrix}
0&0&0&X\\
0&0&X&X\\
0&X&0&X\\
X&X&X&X\\
\end{pmatrix};~\quad
T_6=\begin{pmatrix}
0&0&0&X\\
0&0&0&X\\
0&0&X&X\\
X&X&X&X\\
\end{pmatrix}.
\end{eqnarray}
In this $4\times4$ active sterile mixing matrix, there are six mixing angles, six mass squared differences and four mass eigenvalues. For simplicity, we have fixed Dirac and Majorana phases. We use the latest $3\sigma$ global fit results for active neutrino parameters and equate zeros with the corresponding matrix obtained from the diagonalizing leptonic mixing matrix  $U_{PMNS}^{4\times4}$ matrix.  $m_{\nu}=U m^{\text{diag}}(m_1,m_2,m_3,m_4)U^T$. All the six structures were analyzed and it was found that only $T_5$ was able to satisfy the current bounds on active-sterile mixing parameters \cite{Abe:2016nxk,Ade:2015xua}. This result on five zero textures agrees with past work \cite{Borah:2017azf}, hence we skip the part of analyzing oscillation parameters in the active neutrino sector here and focused to dig on some other phenomenological aspects.
\subsection{Model Description}
Discrete flavor symmetries like $A_4,S_4$~\cite{Babu:2009fd, Ma:2009wi, Altarelli:2005yp, Mukherjee:2017pzq} along with $Z_n$ (n$\ge2$ is always an integer) are intrinsic part of model building in particle physics. In this set-up we basically rely on $A_4$ flavor and unwanted interactions were restricted using extra $Z_3\times Z_4$ symmetry.
$A_4$ being the discrete symmetry group of rotation with a tetrahedron invariant, it consists of 12 elements and 4 irreducible representations denoted by $\bf{1},\bf{1^{\prime}},\bf{1^{\prime\prime}} $ and $\bf{3}$. The particle contents of this model are given in Table.~\ref{modelt}. { We have added three right-handed (RH) heavy fermions ($N_1,N_2,N_3$) and a light fermion ($S$). Along with the additional Higgs scalar doublet ($\phi_2$), we have added another four scalars ($\psi_{1,2},\eta$ and $\chi$) . To be more precise, we would like to explain the purpose of each field as follows. \begin{itemize}
	\item Three heavy fermions ($N_1,N_2$ and $N_3$) and associated scalar fields ($\psi_1, \psi_2$ and $\eta$) were considered to explain neutrino mass and resonant leptogenesis. Within the MES framework two out of three active neutrinos and a sterile neutrino are massive. Hence to address three massive neutrinos, we need three RH neutrinos ($N_1,N_2$ and $N_3$).
	\item The additional Higgs doublet ($\phi_2$) was considered to achieve a maximum of five-zeros in the final $4\times4$ neutrino mass matrix. This Higgs doublet gets tiny VEV ($\sim0.1$eV)via soft breaking mass term \cite{Davidson:2009ha}.
 	\item  Finally, along with three active neutrinos, we are interested in a single generation of sterile neutrino, therefore the scalar field $\chi$ and the fermionic field $S$ come into the picture. Interestingly, the scalar field $\chi$ is a complex scalar, whose real component is associated with the fermion sector and the complex component is behaving as a potential dark matter candidate in our study. This scalar field is behaving as the bridge between these two sectors (neutrino and dark matter).
\end{itemize}}In the following sub-sections, we carefully discuss the scalar and fermion sectors of this model separately in a detailed manner.
\begin{table}[h!]
	\begin{tabular}{|c|cccccc|cccccc|}
		\hline
		Fields$\rightarrow$ &&&&Fermions&&&&Scalars&&&&\\
		\hline
		Charges$\downarrow$		& $L=(\nu_l \,\, l)^T $&$l_R$&$N_1$&$N_2$&$N_3$&$S$ &$\phi_1$&$\phi_2$&$\psi_1$&$\psi_2$&$\eta$&$\chi$\\
		\hline		
		$SU(2)$&2&1&1&1&1&1&2&2&1&1&1&1\\
		$A_4$&3&$(1,1^{\prime}, 1^{\prime\prime})$&1&$1$&$1$&$1^{\prime}$&1&1&3&3&1&$1^{\prime\prime}$\\
		$Z_3$&$\omega^2$&$\omega$&$\omega$&$\omega$&$\omega$&$\omega^2$&$\omega$&$\omega$&$\omega^2$&$\omega^2$&$\omega$&$1$\\
		$Z_4$&1&$i$&$-i$&$i$&$-i$&$-i$&$-i$&$i$&1&$1$&$i$&$-1$\\
		\hline
	\end{tabular}
	\caption{Particle content and their charge assignments under SU(2), $A_4$ and $Z_3\times Z_4$ groups.}\label{modelt}
\end{table}

\subsubsection{\bf The scalars}
{ We denote} the SM Higgs doublet as $\phi_1$ which transform as a singlet under $A_4$ symmetry and it is expressed as, $\phi_1=\frac{1}{\sqrt{2}}\begin{pmatrix}
	\phi_h^+,\,\phi_{h} + v_h+i \phi_h^0
\end{pmatrix}^T$, where $v_h$ is the vacuum expectation value (VEV) of the real part of the scalar doublet. 
Along with the SM Higgs boson, we have introduced a similar doublet $\phi_2=\frac{1}{\sqrt{2}}\begin{pmatrix}
	\phi_{2h}^+,\,\phi_{2h} + v_{2h}+i \phi_{2h}^0
\end{pmatrix}^T$ , which gets a tiny VEV ($v_{2h}\sim0.1$ eV) via soft breaking mass term~\cite{Davidson:2009ha}. We have four more $SU(2)$ scalar singlet $\psi_1$, $\psi_2$, $\eta$ and $\chi$. Here, $\chi$ is a $SU(2)$ complex singlet and it is expressed as $\chi=(\chi^R+v_{\chi}+i\chi^I)/\sqrt{2}$. 
The $A_4$ triplet scalar flavons ($\psi_1$ and $\psi_2$) get VEV\footnote{We follow the similar approach for evaluating triplet flavon VEVs by potential minimization \cite{Das:2018qyt}} along $\langle\psi_1\rangle=(v,0,0)$ and $\langle\psi_2\rangle=(0,v,0)$. The singlets ($\eta,\chi$) achieve VEV along $\langle\eta\rangle=u$ and $\langle\chi\rangle=v_{\chi}$.
The VEV of these scalar fields breaks the $A_4$ including the $Z_3$ and $Z_4$ symmetries. This will help to understand the fermions (quarks, charged leptons and neutrino) mass matrix~\cite{Babu:2009fd, Ma:2009wi, Altarelli:2005yp}.
We consider the value of the VEVs and the other parameters in such a way that the scalar potential related to $ \phi_2,\psi_1,\psi_2$ and $\eta$ fields are completely decoupled from all the other scalar fields; however, { these scalars} play a crucial role to get the fermion mass spectrum. { A detailed discussion} on the scalar potential has been carried out in the appendix section \ref{apotential}.
The VEVs are considered as $v\sim 10^{12}$ GeV, $u\sim10^{10}$ GeV and $v_{\chi}\sim 1000$ GeV.
The scalar potential for the complex singlet ($\chi$) and Higgs doublet ($\phi_1$) can now be written as,
\begin{eqnarray}
	V &= &-\mu_{\phi_1}^2 \phi_1^{\dagger}\phi_1 + \lambda_1 (\phi_1^{\dagger}\phi_1)^2 + \lambda_2 (\phi_1^{\dagger}\phi_1) (\chi^{\dagger} \chi ) \nn \\
	&&- \frac{1}{2} \mu_{\chi^R}^2  \chi^{\dagger} \chi - \frac{1}{2}  \mu_{\chi^I}^2 (\chi^2 + h.c) + 
	\frac{\lambda_3}{4!} (\chi^{\dagger} \chi )^2.
	\label{pot1}
\end{eqnarray}
The minimization condition can be written as,
\begin{equation}
	\mu_{\phi_1}^2=  \lambda_1 v_h^2 + \frac{1}{2} \lambda_2 v_{\chi}^2,  ~~{\rm and }~~\mu_{\chi^R}^2=  -2\mu_{\chi^I}^2+ \lambda_2 v_h^2 + \frac{1}{12} \lambda_3 v_{\chi}^2.
	\label{eq:min}
\end{equation}
The complex part of the singlet scalar $\chi$ does not mix with $\phi_h^0$ due to the charge assignment ($Z_3\times Z_4$ symmetry remains intact), hence neutral Goldstone boson takes the form $ G^0\approx \phi_h^0$ while the charged Goldstone bosons can be written as $G^\pm\approx \phi^\pm$. These components are eaten by corresponding gauge bosons. Only the cp-even, {\it i.e.}, the real component of { this scalar} gets mixed and the  mass matrix could be written as,
\begin{equation}
	\begin{pmatrix}
		\phi_h & \chi^R\\
	\end{pmatrix}\begin{pmatrix}
		M_{\phi_h\phi_h}&M_{\phi_h \chi }\\M_{\chi\phi_h}&M_{\chi\chi}
	\end{pmatrix}\begin{pmatrix}
		\phi_h \\ \chi^R\\
	\end{pmatrix}\label{mm1}
\end{equation}

The mass terms are expressed as,
\begin{eqnarray}\label{mass1}
	&M^2_{\phi_h\phi_h} =&2 \lambda_1 v_h^2,\\
	& M^2_{\phi_h \chi }=&M^2_{\chi\phi_h} = \lambda_2 v_h v_{\chi},\\
	& M^2_{\chi\chi} =&  \frac{1}{12} \lambda_3 v_{\chi}^2.
\end{eqnarray}

The mass eigenstates are obtained by diagonalizing the mass matrix \eqref{mm1} with a rotation of $\phi_h-\chi^R$ basis,
\begin{equation}
	\begin{pmatrix}
		h\\H\\
	\end{pmatrix}=\begin{pmatrix}
		\cos\alpha&\sin\alpha\\-\sin\alpha&\cos\alpha\\
	\end{pmatrix}\begin{pmatrix}
		\phi_h \\ \chi^R\\
	\end{pmatrix}.
\end{equation} 
Here, the mixing angle $\alpha$ is defined as,
\begin{equation}
	\tan2\alpha=\frac{2 M_{\chi\phi_h}}{M_{\chi\chi}-M_{\phi_h\phi_h}}.
\end{equation}
The mass expression of the complex scalar is given by
\begin{equation} \label{mass2}
	M_{\chi^I}^2 = 2 \mu_{\chi^I}^2
\end{equation}

One can notice that the mass of the complex scalar directly depends on $\mu_{\chi^I}$, which is also related to the mixing angle $\alpha$. Hence, we have to be very careful about the parameter space; else it will directly affect the Higgs signal strength data \cite{Sirunyan:2018ouh}.
In this model, the CP-even scalar breaks the $Z_3,Z_4$ including $A_4$ symmetry. However, these symmetries for the singlet type pseudo { scalar remain intact}. Hence, this scalar remains stable and can serve as a viable dark matter candidate. One can see from equation \eqref{mass1}-\eqref{mass2} that the mass of the dark matter and the Higgs portal $\chi^I\chi^Ih(H)$ coupling strengths (the main annihilation channels) depend on the parameters $\mu_{\chi^I}$ and other quartic couplings especially $\lambda_3$ and VEVs, on the other hand, masses of the scalar fields ($h,H,\chi^I$) and mixing angle $\alpha$, VEVs. We will discuss it in detail in the dark matter section.  These parameters also affect the generation of fermionic mass and mixing angles.

\subsubsection{\bf The fermions}
Three generations of heavy right-handed (RH) neutrinos are also introduced, which gives respective masses to the active and sterile neutrinos\footnote{Within MES one active neutrino mass is always zero \cite{Zhang:2011vh,Barry:2011wb,Das:2018qyt}, in the NH mass ordering, $m_1$ will be zero. Hence two RH neutrinos will give mass to two active neutrinos, and one will give mass to a sterile mass.}. We have considered a nearly degenerate mass scale for two RH neutrinos ($N_1, N_3$)\footnote{Choice of the RH neutrino is arbitrary. We choose $N_3$ with $N_1$ instead of $N_2$ for the resonant production of CP asymmetry is due to the choice of a sterile basis within the model. }, in such a way that they can exhibit resonantly entrenched leptogenesis in this work. 

The invariant Lagrangian can be written as,
\begin{eqnarray}\label{lag2}
	\mathcal{L}\supset \mathcal{L_{M_D}}+\mathcal{L_{M_R}}+\mathcal{L_{M_S}},
\end{eqnarray}
where,
\begin{equation}\label{lag1}
	\begin{split}
		\mathcal{L_{M_D}}=&\frac{Y_1}{\Lambda}(L\tilde{\phi_2}\psi_1)_{1}N_1+\frac{Y_2}{\Lambda}(L\tilde{\phi_2}\psi_2)_{1}N_1+\frac{Y_3}{\Lambda}(L\tilde{\phi_1}\psi_1)_{1}N_2\\&+\frac{Y_4}{\Lambda}(L\tilde{\phi_1}\psi_2)_{1}N_2+\frac{Y_5}{\Lambda}(L\tilde{\phi_2}\psi_2)_{1}N_3+\frac{Y_6}{\Lambda}(L\tilde{\phi_2}\psi_1)_{1}N_3,\\
		\mathcal{L_{M_R}}=&\kappa_1\eta\overline{N_1^c}N_3+\kappa_2\eta\overline{N_2^c}N_2+\kappa_3\eta\overline{N_3^c}N_1,\\
		\mathcal{L_{M_S}}=&q_1\chi SN_1+q_2\chi SN_3.
	\end{split}
\end{equation}
{ For convenience, we have considered diagonal charged lepton mass basis throughout the analysis.} We have used non-renormalizable terms in the charged lepton and Dirac mass terms which are suppressed by the mass dimensional parameter $\frac{1}{\Lambda}$.
The renormalizable and non-renormalizable terms in the Lagrangian are classified
based on the analysis of ultraviolet(UV)-divergences in the quantum field theories.
For the renormalizable terms, UV-divergence eliminating local counter-terms repeat the structure of the original Lagrangian. Hence, it can be absorbed into the renormalization of the corresponding terms in the Lagrangian at any level.
On the other hand, in the non-renormalizable case, in each order
of perturbation, new structures appear that do not repeat the original ones. Hence, there is an infinite number of terms, it will not cancel the divergences. Normally, the non-renormalizable situation is usually
considered an unacceptable theory. However, 
it was suggested in Ref.~\cite{Kazakov:2019wce, Kazakov:2020xbo} that the divergence terms elimination methods should be equally applicable to both renormalization and non-renormalizable theories.
The new divergences in each level of perturbation
could still be eliminated
by the introduction of local counter-terms using the Bogolyubov-Parasyuk-Hepp-Zimmermann (BPHZ) $\mathcal{R}$-operation~\cite{Hepp:1966eg,Zimmermann:1969jj}. 
It is noted that the counter-terms are proportional to the powers of momenta of the fields. Here we do not delve into details of the calculations.
This mass-scale $\Lambda$ is close to the GUT scale. 
Typically, in a model building study, $0.004<\frac{\langle\psi\rangle}{\Lambda}<1$~\cite{Altarelli:2005yp} and the Yukawa couplings of $\mathcal{O}(10^{-2}-1)$~\cite{Abada:2007ux} must obey, so that it can satisfy the neutrino mass bounds obtained from recent experiments~\cite{Aghanim:2018eyx}.
One can also bring the mass scale up to the TeV scale, but this required the Yukawa couplings of the order less than $\mathcal{O}(10^{-10})$.
Anyway, those non-renormalizable interactions in the sterile neutrino sector are avoided. One may add terms like $\frac{1}{\Lambda}\psi\psi SS$ in the Lagrangian, however, those terms may deface the current MES structure. After the symmetry breaking, those non-renormalizable terms lead to a larger mass term for the singlet fermion $S$, which is unacceptable for the MES scenario. Therefore, we have restricted those non-renormalizable interactions by the discrete charges in our model.
The mass matrices generated will be of the form as follows,
\begin{itemize}
	\item {\bf Dirac mass matrix:} $M_D=\begin{pmatrix}
		a&0&x\\b&0&0\\z&y&c\\
	\end{pmatrix}$, with $a=\frac{Y_1\langle\phi_2\rangle v}{\Lambda}$, $b=\frac{Y_3\langle\phi_1\rangle v}{\Lambda}$, $c=\frac{Y_5\langle\phi_2\rangle v}{\Lambda}$, $x=\frac{Y_2\langle\phi_2\rangle v}{\Lambda}$, $y=\frac{Y_4\langle\phi_1\rangle v}{\Lambda}$, $z=\frac{Y_6\langle\phi_2\rangle v}{\Lambda}$.
	\item {\bf Majorana mass matrix:} $M_R=\begin{pmatrix}
		0&0&d\\0&e&0\\f&0&0\\
	\end{pmatrix}$, with $d\sim e\sim f=\kappa_iu$ ($i=1,2,3$ for respective positions).
	\item {\bf Sterile mass matrix:} $M_S=\begin{pmatrix}
		g&0&h\\
	\end{pmatrix}$ with $g=q_1v_{\chi}$ and $h=q_2v_{\chi}$. To get the sterile mass within eV scale for $v_{\chi}\simeq 1$ TeV, the Yukawa couplings $q_1,q_2$ take value around $10^{-3}$.
	\item The final active-sterile mass matrix takes the form,
	\begin{eqnarray}\label{44m}
		M_{\nu}&&=\begin{pmatrix}
			M_DM_R^{-1}M_D^T&M_DM_R^{-1}M_S^T\\
			M_S{M_R^{-1}}^TM_D&M_SM_R^{-1}M_S^T\\
		\end{pmatrix}_{4\times4} =\begin{pmatrix}
			\frac{ax(d+f)}{df}&\frac{ax}{d}&\frac{ac+xz}{d}&\frac{ah}{f}+\frac{gx}{d}\\
			\frac{ax}{f}&0&\frac{bc}{f}&\frac{bh}{f}\\
			\frac{ac+xz}{f}&\frac{bc}{d}&\frac{cz(d+f)}{df}&\frac{cg}{d}+\frac{hy}{f}\\
			\frac{ah}{f}+\frac{gx}{d}&\frac{bh}{f}&\frac{cg}{d}+\frac{hy}{f}&\frac{(d+f)gh}{df}\\
		\end{pmatrix}_{4\times4}
	\end{eqnarray}
	Due to the choice of small VEV to the additional Higgs doublet $\phi_2$ ($\sim$ eV), the (11), (12), (21), (13),(31) and (33) position in the $M_{\nu}$ mass matrix are much smaller compared to the other elements (as $a,c,x,z\ll b$) and they lead to $5-Zero$ texture in the active-sterile mass matrix as,
	\begin{eqnarray}
		M_{\nu}\simeq\begin{pmatrix}
			0&0&0&\frac{ah}{f}+\frac{gx}{d}\\
			0&0&\frac{bc}{f}&\frac{bh}{f}\\
			0&\frac{bc}{d}&0&\frac{cg}{d}+\frac{hy}{f}\\
			\frac{ah}{f}+\frac{gx}{d}&\frac{bh}{f}&\frac{cg}{d}+\frac{hy}{f}&\frac{(d+f)gh}{df}\\
		\end{pmatrix}_{4\times4}
	\end{eqnarray}
\end{itemize}
It is to be noted that the dark matter $\chi^I$ could not decay into $S$ and $N_1$ through the $\chi SN_1$ operators in equation \eqref{lag1}, as $M_{N_1}> M_{\chi^I}$. However, it could decay into two active neutrinos through mixing. We checked that the effective coupling strength for $\chi^I\nu_i\nu_j$ $(i,j=1,2,3)$ are very very less than $10^{-22}$, which implies that the dark matter decay lifetime is much greater than the lifetime of the Universe { ($t_{universe}\sim10^{17}$ sec)}. Hence, in this model dark matter is stable, even we found the DM lifetime, $t_{DM}>10^{28}$ sec \cite{Boyarsky:2014jta}.

\section{Constraints on this model}\label{sec3}
There are various kinds of theoretical constraints in the model by which the parameter space of this model is constrained. In the following subsections, we will be discussing a few of them that we used in this model.

\subsection{Stability constraints}
The stability constraints demand that the potential should be bounded from below, {\it i.e.}, it should not go negative infinity along any direction of the field space at large field values. For large fields the quadratic terms of
the scalar potential in Eqn. \eqref{pot1} are very small compared to the quartic terms,
hence, the scalar potential can be written as,
\begin{equation}
	V(H, \chi^R)=-\frac{1}{4}\lambda_1H^4+\frac{\lambda_2}{4}H^2(\chi^R)^2+\frac{1}{24}\lambda_3 (\chi^R)^4.
\end{equation}
With further simplification, it will take the form,
\begin{equation}
	V(H,\chi^R)=\frac{1}{4}\Big[\sqrt{\lambda_1}H^2+\frac{\sqrt{\lambda_3}}{\sqrt{6}}(\chi^R)^2\Big]^2+\frac{1}{4}\Big[\lambda_2+\sqrt{\frac{2\lambda_1\lambda_3}{3}}\Big]H^2(\chi^R)^2.
\end{equation}
This scalar potential will be bounded from below if the following conditions are satisfied \cite{Deshpande:1977rw},
\begin{equation*}
	\lambda_1(\Lambda)>0;~~~\lambda_3(\Lambda)>0;~~~\text{and}~~\lambda_2(\Lambda)+\sqrt{\frac{2\lambda_1(\Lambda)\lambda_3(\Lambda)}{3}}>0.
\end{equation*}
Here, the coupling constants are evaluated at a scale $\Lambda$ using RG equations. 
\subsection{Perturbativity constraints}
This model, to remain perturbative at any given energy scale, one must impose upper bound on the coupling constants of the potential of \eqref{pot1} and they are as follows \cite{Lee:1977eg},
\begin{equation}
	|\lambda_1(\Lambda),~\lambda_2(\Lambda),~\lambda_3(\Lambda)|\le4\pi.
\end{equation}
\subsection{Unitarity constraints}
The scalar potential parameters of this model are severely constrained by the
unitarity of the scattering matrix (S-matrix), which consists of the quartic couplings of the scalar potential. For large field values, the scattering
matrix is obtained by using various scalar-scalar, gauge boson-gauge boson, and scalar-gauge
boson scatterings \cite{Lee:1977eg}. Following the unitarity condition, the S-matrix elements demand that the eigenvalues of the scattering matrix should be less than $8\pi$ \cite{Das:2014fea}. The unitary bounds in this model are,
\begin{equation}
	\lambda_1\le8\pi~~~\text{and}~~|12\lambda_1+\lambda_3\pm\sqrt{16\lambda_2^2+(\lambda_3-12\lambda_1)^2}|\le32\pi.
\end{equation}

\subsection{Bounds from Higgs Signal strength data}

At the tree-level, the couplings of Higgs-like scalar $h$ to the fermions and gauge bosons in the presence of extra Higgs doublet ($\phi_2$) and singlet scalar ($\chi$) are modified due to the mixing. Loop induced decays will also have slight modification for the same reason; hence, new contributions will be added to the signal strength \cite{Sirunyan:2018ouh}. As a particular case, we consider
the Higgs boson production cross-section via the gluon fusion mechanism and we use the narrow width
approximation  $\Gamma_h/M_h\rightarrow0$,
\begin{equation}
	\mu_{X}=\frac{\sigma(gg\rightarrow h\rightarrow X)_{BSM}}{\sigma(gg\rightarrow h\rightarrow X)_{SM}}\approx\frac{\sigma(gg\rightarrow h)_{BSM}}{\sigma(gg\rightarrow h)_{SM}}\frac{Br(h\rightarrow X)_{BSM}}{Br(h\rightarrow X)_{SM}},
\end{equation}
where, $X=b\bar{b}, \tau^+ \tau^-,\mu^+ \mu^-, W W^*, Z Z^*, \gamma\gamma$ is the standard model particle pairs. In presence of an extra Higgs doublet $\phi_2$, the signal strength does not change as it completely decoupled from the scalar sector, however, due to the mixing of $\phi_1$ and $\chi$, $h$ to flavon-flavon ( or boson-boson) coupling become proportional to $\cos\alpha$. So, we may rewrite $\mu_{X}$ as,
\begin{equation}
	\mu_{X}=\cos^2\alpha \frac{\Gamma(h\rightarrow X)_{BSM}}{\Gamma(h\rightarrow X)_{SM}}\frac{\Gamma^{total}_{h,SM}}{\Gamma^{total}_{h,BSM}}.
\end{equation}
Apart from the SM Higgs $h$, if the masses for the extra physical Higgses are greater than $M_h/2$, $\frac{\Gamma^{total}_{h,SM}}{\Gamma^{total}_{h,BSM}}\approx\big(\cos^2\alpha \big)^{-1}$. Hence, the modified signal strength will be written as,
\begin{equation}
	\mu_{X}=\frac{\Gamma(h\rightarrow X)_{BSM}}{\Gamma(h\rightarrow X)_{SM}}.
\end{equation}
In this study, we notice that $\mu_{Z}$ is the most stringent, bounding $0.945\lesssim \cos\alpha\lesssim 1$ depending on singlet scalar VEVs, although with little sensitivity. $\mu_{W}$ is less sensitive, yet bounding $0.92\lesssim \cos\alpha\lesssim 1$. In this work, we keep fixed $\cos\alpha=0.95$ throughout the calculations.
\subsection{MES and Yukawa parametrization}
Minimal extended seesaw(MES) is used in this work to study active and sterile masses. The advantage of working under the MES framework is that it let us study sterile neutrino with a broader range of the mass spectrum (eV to keV). In this MES framework, three right-handed neutrinos, and one additional gauge singlet chiral field, $S$ are introduced. The Lagrangian for MES of the neutrino mass terms is given by \cite{Barry:2011wb},
\begin{equation}\label{mes1}
-\mathcal{L}_{\mathcal{M}}= \overline{\nu_{L}}M_{D}\nu_{R}+\frac{1}{2}\overline{\nu^{c}_{R}}M_{R}\nu_{R}+\overline{S^c}M_{S}\nu_{R}+h.c. .
\end{equation}  
Here, $M_D$ and $M_R$ are $3\times3$ Dirac and Majorana mass matrices respectively whereas $M_S$ is a $1\times3$ matrix. 
We use the $4\times4$ unitary PMNS matrix to diagonalize the neutrino mass matrix. This contains six mixing angles ($\theta_{12},\theta_{13},\theta_{23},\theta_{14},\theta_{24},\theta_{34}$), three Dirac CP phases ($\delta_{13},\delta_{14},\delta_{24}$) and three Majorana phases ($\alpha,\beta,\gamma$)\footnote{For simplicity, we keep only one phase ($\delta_{13}=\delta$) and others are set to zero.}. The parametrization of the PMNS matrix can be expressed as \cite{Gariazzo:2015rra},
\begin{equation}
	U_{PMNS}=R_{43}\tilde{R_{24}}\tilde{R_{14}}R_{23}\tilde{R_{13}}R_{12}P,
\end{equation}
where $R{ij},\tilde{R{ij}}$ are the rotational matrices with rotation in respective planes and $P$ is the diagonal Majorana phase matrix, which are given by,
\begin{eqnarray}
	\nn
	R_{34}=\begin{pmatrix}
		1&0&0&0\\
		0&1&0&0\\
		0&0&c_{34}&s_{34}\\
		0&0&-s_{34}&c_{34}\\
	\end{pmatrix},~\tilde{R_{24}}=\begin{pmatrix}
		1&0&0&0\\
		0&c_{24}&0&s_{24}e^{i\delta_{24}}\\
		0&0&1&0\\
		0&-s_{24}e^{i\delta_{24}}&0&c_{24}\\
	\end{pmatrix},\\P=\begin{pmatrix}
		1&0&0&0\\
		0&e^{i\alpha/2}&0&0\\
		0&0&e^{i(\beta/2-\delta_{13})}&0\\
		0&0&0&e^{-i(\gamma/2-\delta_{14})}\\
	\end{pmatrix}.
\end{eqnarray}
The abbreviations used are as $c_{ij}=\cos \theta{ij}$ and $s_{ij}=\sin \theta_{ij}$. Now, the $4\times4$ low scale neutrino mass matrix is expressed as,
\begin{equation}\label{nu4}
	m_{\nu}=U_{PMNS} m^D_{\nu}U_{PMNS}^T.
\end{equation}

The most important elements in this active-sterile unitary matrix are the elements in the fourth column. They are expressed as follows \cite{Hagstotz:2020ukm},
\begin{eqnarray}
	&|U_{e4}|^2=&\sin^2\theta_{14},\\
	&|U_{\mu4}|^2=&\cos^2\theta_{14}\sin^2\theta_{24},\\
	&|U_{\tau4}|^2=&\cos^2\theta_{14}\cos^2\theta_{24}\sin^2\theta_{34},\\
	&|U_{s4}|^2=&\cos^2\theta_{14}\cos^2\theta_{24}\cos^2\theta_{34}.
\end{eqnarray}
The active-sterile mixing angles $\theta_{i4}$ are essentially small, such that they do not influence the three neutrino mixing, which is allowed by current global fit limits. In principle, $|U_{s4}|^2$ be of the order of unity and other elements are expected to be very small. Hence, we can refer to the fourth neutrino mass eigenstate as the sterile neutrino itself {\it i.e.,} $m_{4}\sim m_S$.

The diagonalization of the Majorana mass matrix, $M_R$, is given by
\begin{equation}
	M_R^D=V_DM_RV_D^T.
\end{equation}
Without loss of generality, we exclude the presence of sterile neutrino for baryogenesis study and use the conventional $3\times3$ matrix structures to parametrize the Yukawa matrix analogous to the CI parametrization \cite{Casas:2001sr} as,
\begin{equation}
	Y=\frac{1}{v}U_{PMNS}^{3\times3}\sqrt{m^D_{\nu}}R^T\sqrt{M_R^{-1}},\label{ci}
\end{equation}
where $R$ is the rotational matrix with complex angle $z_i=x_i+iy_i$ ($x,y$ being real and free parameter with $x_i,y_i\in[0,2\pi]$ \cite{Ibarra:2003up}.), and it can be parametrized as,
\begin{equation}
	R=\begin{pmatrix}
		1&0&0\\
		0&\cos z_1&\sin z_1\\
		0&-\sin z_1&\cos z_1\\
	\end{pmatrix}\begin{pmatrix}
		\cos z_2&0&\sin z_2\\
		0&1&0\\
		-\sin z_2&0&\cos z_2\\
	\end{pmatrix}\begin{pmatrix}
		\cos z_3&\sin z_3&0\\
		-\sin z_3&\cos z_3&0\\
		0&0&1\\
	\end{pmatrix}
\end{equation}
For our convenience, we have considered some random value (say $z_1=0^\circ-i90^\circ $) for the complex angles $z_i$ satisfying current value for the observed quantities. 
\subsection{Baryogenesis via resonant leptogenesis}
In the resonant leptogenesis study, we numerically solved the Boltzmann equations (BE) in the out-of-equilibrium scenario. With our choice of nearly degenerate mass spectrum of TeV scaled RH neutrinos, the decay rates can produce sufficient asymmetry and at a temperature above the electroweak phase transition (EWPT), {\it i.e.}, for $T\ge T_c\sim147$ \cite{Dev:2014laa} GeV, $(B+L)$-violating interactions mediated by sphalerons are in thermal equilibrium, hence, the asymmetries in baryon number are generated. For $T<T_c$ , the produced baryon asymmetry gets diluted by photon interactions until the recombination epoch at temperature $T_0$. If there is no significant entropy release mechanism while the Universe is cooling down from $T_c$
to $T_0$, the baryon number in a co-moving volume, $\eta_B/s$, is constant during this epoch.

Before directly involving the BEs, let us first draw attention to a few quantities that are involved in the equations. The washout parameter, $K$, can be defined as the ratio of decay width to the Hubble parameter. Mathematically, $K=\Gamma_{N_i}/H(T=M_{N_i})$, with $\Gamma_{N_i}=\frac{(Y^{\dagger}Y)_{ii}}{8\pi}M_{N_i}$ and $H(T)=1.66 g_*\frac{T^2}{M_{Planck}}$. Here $M_{Planck}=1.12\times10^{19}$ GeV is the Planck mass, and $g_*$ is the relativistic degree of freedom with its value lying around 106. A small $K$ value corresponds to weak washout, and a large $K$ value corresponds to strong washout effects. In the case when flavours are taken under consideration \cite{Pilaftsis:2003gt},
\begin{equation}
	K_i=\frac{\Gamma (N_1\rightarrow l_i H)}{H(T=M_{N_1})}\simeq \frac{\Gamma (N_3\rightarrow l_i H)}{H(T=M_{N_3})} \quad\text{and}\quad K=\sum_{j=e,\mu,\tau}K_j.
\end{equation}
The relations among the number densities to the entropy density ($Y_x=\frac{N_x}{s}$) are expressed as follows \cite{Pilaftsis:2003gt},
\begin{equation}
	Y_B(T>T_c)=\frac{28}{79}Y_{B-L}(T>T_c)=-\frac{28}{51}Y_L(T>T_c).
\end{equation}
The CP asymmetry due to the heavy neutrino mixing in the flavoured case is given by, \cite{Dev:2014laa, Bambhaniya:2016rbb},
\begin{eqnarray}
	\epsilon_{il}&=&\epsilon_{il}^{mix}+\epsilon_{il}^{osc}.\\
	&=&\sum_{i\ne j}^{3}\frac{\text{Im}\big[Y_{il}Y_{jl}^*(YY^{\dagger})_{ij}\big]+\frac{M_{i}}{M_{j}}\text{Im}\big[Y_{il}Y_{jl}^*(YY^{\dagger})_{ji}\big]}{(YY^{\dagger})_{ii}(YY^{\dagger})_{jj}}\big(f_{ij}^{mix}+f_{ij}^{osc}\big).
\end{eqnarray}
The terms with superscripts inside the braces represent mixing and oscillation contributions respectively and they are expresses as,
\begin{eqnarray}
	f_{ij}^{mix}&=&\frac{(M_{i}^2-M_{j}^2)M_{i}\Gamma_j}{(M_{i}^2-M_{j}^2)^2+M_{i}^2\Gamma_j^2},\\
	f_{ij}^{osc}&=&\frac{(M_{i}^2-M_{j}^2)M_{i}\Gamma_j}{(M_{i}^2-M_{j}^2)^2+(M_{i}\Gamma_i+M_{j}\Gamma_j)^2\frac{|\text{Re}(YY^{\dagger})|}{(YY^{\dagger})_{ii}(YY^{\dagger})_{jj}}}
\end{eqnarray}
where, $\Gamma_i=\frac{M_{i}}{8\pi}(YY^{\dagger})_{ii}$ is the decay width at tree level. The typical time scale for the CP asymmetry variation is defined as,
\begin{equation}
	t=\frac{1}{2H}=\frac{z^2}{2H(M_1)}=\frac{Kz^2}{2\Gamma_{N_1}}\sim\frac{1}{\Delta M}.
\end{equation}
The CP asymmetry raises for $t\le\frac{1}{\Delta M}$ and exhibits its oscillation pattern only for $t\ge\frac{1}{\Delta M}$. They originate from the CP-violating decays of the two mixed states $N_1$ and $N_3$. The time dependence if the CP asymmetry is also neglected as in the strong washout regime, their contribution is very small. 

The baryon to photon number density, ($\eta_B=\frac{N_B}{n_{\gamma}}$) in the RL scenario is also defined as,
\begin{equation}
	\eta_B\sim-\sum_{i}^{1,2,3} \frac{\epsilon_i}{200 K(z)}.
\end{equation}
One thing to note here, to achieve $\eta_B\sim6.1\times10^{-10}$ \cite{Davidson:2008bu}, the term $\frac{\epsilon_i}{K(z)}$ should be around $10^{-8}$. Hence, the resonant production of leptonic asymmetry $\mathcal{O}(1)$ is possible, only when a strong wash-out region is there ($K\gg1$). This leads to a thermally dense plasma state, so the conditions required for kinetic equilibrium and decoherence of the heavy Majorana neutrinos in the BEs are comfortably satisfied.

We have considered only $1\leftrightarrow2$ decays of the RH neutrinos and the $2\leftrightarrow2$ scatterings that describe $\Delta L=0,2$ transition processes \cite{Pilaftsis:2003gt}. 
Moreover, the decay rates of the two RH neutrinos are almost equal due to the nearly degenerate mass scheme ($\Gamma_{N_1}\sim\Gamma_{N_3}\sim\Gamma$) hence, the CP asymmetries generated from both the decay processes get resonantly enhanced. In the case of different decay rates, it would be wise to pick the only CP asymmetry contribution, which is resonantly enhanced. 
We also have included the flavour effect in our RL study, where the contribution from specific lepton flavours is accounted for. As leptogenesis is a dynamic process, the lepton asymmetry generated via the decay and inverse decay of RH neutrinos is distributed among all three flavours. In some cases, the flavour dependent processes do wash out the lepton asymmetry via inverse decay processes. Finally, the total lepton and the baryon asymmetry values are given by the sum of all three contributions. 
Hence, the Boltzmann equation for three flavor case can be written as \cite{Pilaftsis:2003gt},
\begin{eqnarray}
	&\frac{d\eta_{N_j}}{dz}=&\frac{z}{H(z=1)}\Big[\Big(1-\frac{\eta_{N_j}}{\eta_{N_j}^{eq}}\Big)\frac{1}{n_{\gamma}}\gamma_{L\phi}^{N_j}\Big],\\
	\nn&\frac{d\eta_{Li}}{dz}=&\frac{z}{H(z=1)\eta_{\gamma}}\Big[\sum_{i=1}^3\epsilon_i\Big(1-\frac{\eta_{N_j}}{\eta_{N_j}^{eq}}\Big)\epsilon_i\gamma_{L\phi}^{N_j}-\frac{2}{3}\eta_{L_i}\sum_{k=e,\mu,\tau}\Big(
	\gamma^{L_i\phi}_{L_k^c\phi^{\dagger}}+\gamma^{L_i\phi}_{L_k\phi}\Big)\\&&-\frac{2}{3}\sum_{k=e,\mu,\tau}\eta_{L_k}\Big(
	\gamma^{L_k\phi}_{L_k^c\phi^{\dagger}}-\gamma^{L_k\phi}_{L_i\phi})\Big],\label{be1}
\end{eqnarray}
where  $i=e,\mu,\tau$, represents three flavour case and $j=1,2,3$ represents RH neutrino generations. $\eta_{N_J}^{eq}$ is the equilibrium number density and $\gamma_{L\phi}^{N_j}$ is the $1\leftrightarrow2$ collision term. They are defined as follows.
\begin{eqnarray*}
	&\eta_{N_i}^{eq}=g_a\big(M_{N_i}T/2\pi\big)^{3/2} e^{z_i};~~(z_i=\frac{M_{N_i}}{T}),\\
	&\eta_{\gamma}=(2M_{N_i}^3/\pi^2z^3).
\end{eqnarray*}
The collision and scattering terms used in these equations are discussed in the appendix section \ref{appx}. In the numerical section, we present the evolution of RH neutrinos and the lepton number density with flavour effects.
\subsection{Neutrino-less Double Beta Decay ($0\nu\beta\beta$)}\label{ndbd}
For the observed $0\nu\beta\beta$ process at tree-level we have considered the situation where contribution of an additional sterile neutrino is presumed. The effective electron neutrino Majorana mass for the active neutrinos in the $0\nu\beta\beta$ process read as, 
\begin{equation}
	m^{3_{\nu}}_{eff}= m_1|U_{e1}|^2+m_2|U_{e2}|^2+m_3|U_{e3}|^2.\label{n1}
\end{equation}
As only electrons were involved in the double decay process the phase ``effective $electron$ neutrino'' is used in our text. With additional $n_S$ extra sterile fermions states  in the extended SM sector, those extra states will enhance the decay amplitude which corrects the effective mass as \cite{Bene__2005},
\begin{equation}
	m_{eff} = \sum_{i=1}^{3+n_S}U_{ei}^2 \ p^2 \frac{m_i}{p^2-m_i^2},
\end{equation}

where, $U_{ei}$ is the $(3+n_S \times 3+n_S)$ matrix with extra active-sterile mixing elements. As we have considered only one sterile state, hence, the effective electron neutrinos mass is modified as \cite{Barry:2011wb},
\begin{equation}
	m^{3+1}_{eff}= m^{3_{\nu}}_{eff}+m_4|U_{e4}|^2,
\end{equation}
where, $m^{3_{\nu}}_{eff}$ is the SM contribution only from \eqref{n1}, $|U_{e4}|$ the active-sterile mixing element and $m_4$ is the sterile mass.

Many experimental and theoretical progress was made to date and still counting in to validate the decay process with better accuracy. Nevertheless, no concrete shreds of evidence from experiments confirmed to date to prove the neutrinoless double beta decay process. However, improved next-generation experiments are now trying in pursue of more accurate limit \cite{Obara:2017ndb,Artusa:2014lgv,Hartnell:2012qd,Gomez-Cadenas:2013lta,Barabash:2011aa} on the effective mass which might solve the absolute mass problem. Recent results from various experiments put strong bounds on the effective mass $m_{eff}$, some of them are shown in table \ref{teff}. Kam-LAND ZEN Collaboration \cite{KamLAND-Zen:2016pfg} and GERDA \cite{Agostini:2018tnm} which uses Xenon-136 and Germanium-76 nuclei respectively gives the most constrained upper  bound up to 90\% CL with $$m_{eff}< 0.06-0.165 \ \text{eV}.$$ 
\begin{table*}[t]
	\centering
	\begin{tabular*}{\textwidth}{@{\extracolsep{\fill}}|l|r|r|r|}
		\hline
		\multicolumn{1}{|c|}{Experiments (Isotope)}& \multicolumn{1}{c|}{ $|m_{eff}|$ eV}& \multicolumn{1}{c|}{ Half-life (in years)}& \multicolumn{1}{c|}{ Ref.}\\
		\hline
		KamLAND-Zen(800 Kg)(Xe-136)&$0.025-0.08$&$1.9\times10^{25}$(90\%CL)&\cite{KamLAND-Zen:2016pfg}\\
		KamLAND2-Zen(1000 Kg)(Xe-136)&$<0.02$&$1.07\times10^{26}$ (90\%CL)&\cite{KamLAND-Zen:2016pfg}\\
		GERDA Phase II (Ge-76)& $0.09-0.29$&$4.0\times10^{25}$(90\%CL)&\cite{Agostini:2018tnm}\\
		CUORE (Te-130)& $0.051-0.133$&$1.5\times10^{25}$(90\%CL)&\cite{Artusa:2014lgv}\\
		SNO+ (Te-130)& $0.07-0.14$&$\sim10^{26-27}$&\cite{Hartnell:2012qd}\\
		SuperNEMO (Se-84)&$0.05-0.15$&$5.85\times10^{24}$(90\%CL)&\cite{Barabash:2011aa}\\
		AMoRE-II (M0-100)&$0.017-0.03$&$3\times10^{26}$(90\%CL)&\cite{Bhang:2012gn}\\
		EXO-200(4 Year)(Xe-136)& $0.075-0.2$&$1.8\times10^{25} $(90\%CL)&\cite{Tosi:2014zza}\\
		nEXO(5Yr+5Yr w/Ba Tagging)(Xe-136)& $0.005-0.011$&$\sim10^{28}$&\cite{Licciardi:2017oqg}\\
		\hline
	\end{tabular*}
	\caption{Sensitivity on effective mass of a few past and future experiments with half-life in years. }\label{teff}	
\end{table*}
\subsection{Dark matter}
Recent results from various WMAP satellites and cosmological measurements, the relic density of the current Universe measured as $\Omega_{DM} h^2=0.1198\pm0.0012$ \cite{Aghanim:2018eyx}. In this model, the imaginary component of the complex singlet ($\chi^I$ of $\chi$) serves as a dark matter candidate. We have used {\tt Feynrule} \cite{Alloul:2013bka} to construct the model and carried out the numerical calculations using {\tt micrOmega 5.08} \cite{Belanger:2018mqt}. A detailed analysis has been carried out in the numerical analysis section. 

Dark matter can be detected via direct as well as indirect detection experiments. As WIMP dark matter interacts with matters weakly, many experiments are focused on direct detection techniques.  
If WIMPs scatter from the atomic nucleus, then it deposits energy in the detector given by,
\begin{equation}
	E_{deposit} = \frac{1}{2} M_{DM} v^2.
\end{equation}
The energy deposition can also be written as,
\begin{equation}
	E_{deposit} = \frac{\mu^2 v^2}{m_N}(1-\cos\theta).
\end{equation}
In the Earth frame, the mean velocity $v$ of the WIMPs relative to the target nucleus is about 220 km/s, $\mu$ is the reduced mass of the WIMP of mass $M_{DM}$ and the nucleus of mass $m_N$, and $\theta$ is the scattering angle.
As the dark matter is weakly interacting, it may rarely bump into the nucleus of a detector atom and deposit energy which may create a signature at the detector. The amount of energy of a WIMP with mass $M_{DM}=100$ GeV would deposit in the detector is $E_{deposit}\simeq 27$ keV \cite{Aprile:2012nq}.

Presently non-observation of dark matter through direct detections sets a limit on WIMP-nucleon scattering cross-section for a given dark matter mass from experiments XENON~\cite{Aprile:2012nq, Aprile:2016swn}, LUX~\cite{Akerib:2019diq}. Dark matter mass below 10 GeV is ruled out by recent experimental results from DAMA/LIBRA~\cite{Bernabei:2010mq}, CoGeNT~\cite{  Aalseth:2012if}, CDMS~\cite{Agnese:2013rvf} etc. Hence, in our work, we will be focusing on dark matter parameter regions considering these bounds in mind. 

Indirect detection of dark matter techniques are quite different. If the dark matter and its antiparticle are the same,
then they can annihilate to form known standard model particles such as photons ($\gamma$-ray), electrons ($e^-$), positrons ($e^+$) etc.
Various detectors were placed in the Earth's orbits, {\it e.g.}, Fermi Gamma-ray Space Telescope (FGST) \cite{Hooper:2010mq}, Alpha Magnetic Spectrometer (AMS) \cite{Aguilar:2013qda}, PAMELA \cite{Cholis:2008qq} etc., observed the excess of gamma-ray and positron excess.

From the particle physics point of view, the processes like $DM,~DM \rightarrow \gamma\gamma,~ e^+ e^-$ etc. have been used to explain such excess.
They are model-dependent processes. The WIMP dark matter with different mass and coupling can be considered to explain these high energetic gamma-rays excess from the galactic centre and positron excess in the cosmic ray.

\section{Numerical analysis}\label{sec4}
\subsection{Neutrino mixing}
\begin{figure}[h]\centering
	\includegraphics[scale=.35]{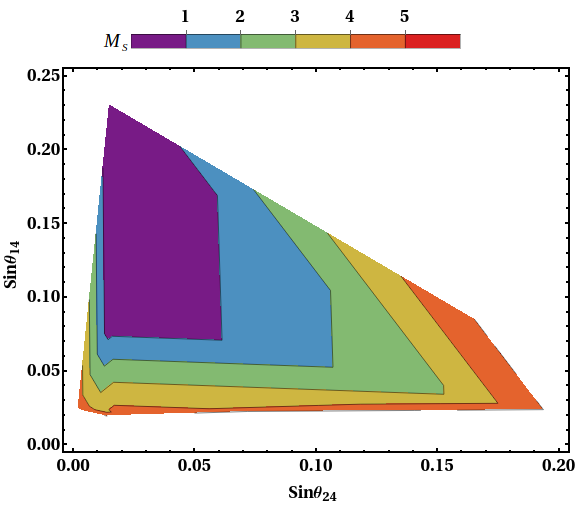}
	\includegraphics[scale=.35]{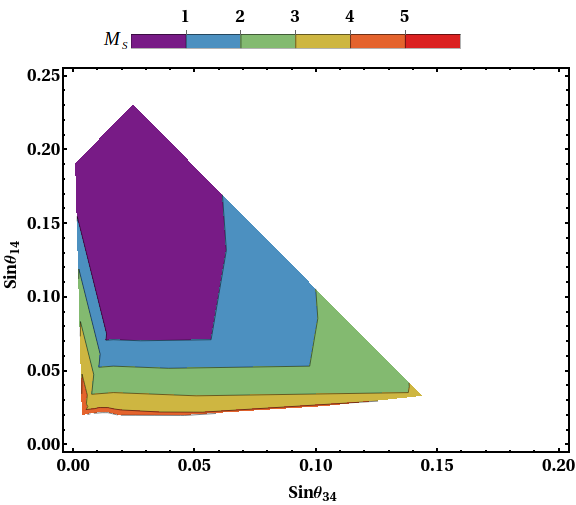}
	\caption{Projection of sterile mass with respect to active-sterile mixing angles using available $3\sigma$ bound on active neutrino parameters \cite{Esteban:2020cvm}. (Left) $\theta_{14}$ and $\theta_{24}$ able to project $m_S$ value around (0.1-3) eV$^2$ for $1.9^{\circ}-11.4^{\circ}$ and $0.35^{\circ}-11.8^{\circ}$ respectively. (Right) Similar results for $\theta_{14}$ and $\theta_{34}$ also satisfy for $m_S$ around (0.1-5) eV$^2$ and $\theta_{34}$ around $0.78^{\circ}-9.79^{\circ}$ and $0.4^{\circ}-6.6^{\circ}$ respectively.}\label{asf}
\end{figure}

In this work, we have skipped the analysis of active neutrinos as there is a vast literature available for this. Rather we focus on active-sterile mixing elements and sterile mass here. Our results show constancy with the previous results of \cite{Borah:2017azf}, and the structure $T_5$ being the only 5-zero texture suitable to study the $3+1$ scenario. We have used the latest global fit 3$\sigma$ bounds on the active neutrino parameters and randomly solved for four unknowns. Four zeros of the structures from equation \eqref{txz} are equated to the $4\times4$ light neutrino mass matrix generated from the diagonalizing matrix given by equation \eqref{44m} and evaluate elements of the fourth column. We work out the numerical analysis with such choice of input parameters, which also satisfy the recent LFV data \cite{Parker_2018}. After solving for the active-sterile parameters, we have shown contour plots in fig.\ref{asf}. Interesting bounds on the mixing matrix and sterile mass are observed from the analysis. In the left figure, we have varied $\sin\theta_{24}$ and $\sin\theta_{14}$ along x-y plane and projected sterile mass $m_4^2$ on the z-plane. Similarly, in the plot next to this is also a projection of sterile mass in $\sin\theta_{34}-\sin\theta_{14}$ plane. {In the left figure, $\theta_{14}$ and $\theta_{24}$ value lie around $1.7^{\circ}-13.29^{\circ}$ and $0.35^{\circ}-10.9^{\circ}$ respectively and they able to project $m_4^2$ value in between $0.1-4$ eV$^2$. On the other hand, the next figure gives bound on $\theta_{14}$ as $1.68^{\circ}-13.4^{\circ}$, however gives a wider range of $m_4^2=0.1-4$ eV$^2$ while satisfying $\theta_{34}$ value in between $0.4^{\circ}-8.5^{\circ}$.}
\begin{table}[h]\centering
	\begin{tabular}{|c|c|c|}
		\hline	Parameters&Experimental result \cite{Borah:2017azf}& Model result\\
		\hline$m_4^2$ (eV$^2$)&$0.7-2.5$&0.1-4\\
		$\theta_{14}$&$>3^{\circ}$&$1.7^{\circ}-13.29^{\circ}$\\
		$\theta_{24}$&$>3^{\circ}$&$0.35^{\circ}-10.9^{\circ}$\\
		$\theta_{34}$&$3.4^{\circ}-11^{\circ}$&$0.4^{\circ}-8.5^{\circ}$\\
		\hline
	\end{tabular}
	\caption{A comparison in-between experimental results and results obtained from the model for the active-sterile mixing parameters. }\label{com1}
\end{table}

\begin{table}[h]\centering
	\begin{tabular}{|c|c|c|c|}
		\hline
		BMPs&Model parameters&Active neutrino parameters&Active-sterile neutrino parameters\\
		\hline
		BMP-1&$\begin{aligned}
			&
		Y_1=0.939,~Y_2=1.11,\\&Y_3=0.38,Y_4=0.078\\&Y_5=0.551,Y_6=0.874\\
		&\langle \phi_1\rangle=246 \text{GeV},\\&\langle \phi_2\rangle\simeq0.01 \text{eV}\\
		&\kappa_1\sim\kappa_2\sim\kappa_3\simeq0.1,\\
		&q_1\sim q_2\simeq 0.002.
		\end{aligned}$&$\begin{aligned}
			&
		\Delta m_{21}^2=7.10\times10^{-5}\text{eV}^2\\&\Delta m_{31}^2=2.49\times10^{-3}\text{eV}^2\\
			& \sin^2\theta_{12}=0.315\\ &\sin^2\theta_{13}=0.0227,\\ &\sin^2\theta_{23}=0.611\\
			&\delta_{13}=3.3~(190^{\circ})\\
		\end{aligned}$&$\begin{aligned}
		&m_4^2=1.915\text{eV}^2\\&\sin^2\theta_{14}=0.0230,\\&\sin^2\theta_{24}=0.0238\\
		&\sin^2\theta_{34}=0.0138\\
	\end{aligned}$\\
\hline
		BMP-2&$\begin{aligned}
	&
	Y_1=0.242,~Y_2=0.29,\\&Y_3=0.099, Y_4=0.116\\&Y_5=0.42, Y_6=0.0589\\
	&\langle \phi_1\rangle=246 \text{GeV},\\&\langle \phi_2\rangle\simeq0.01 \text{eV}\\
	&\kappa_1\sim\kappa_2\sim\kappa_3\simeq0.1,\\
	&q_1\sim q_2\simeq 0.002.
\end{aligned}$&$\begin{aligned}
	&
	\Delta m_{21}^2=8.0\times10^{-5}\text{eV}^2\\&\Delta m_{31}^2=2.59\times10^{-3}\text{eV}^2\\
	& \sin^2\theta_{12}=0.34\\ &\sin^2\theta_{13}=0.024,\\ &\sin^2\theta_{23}=0.421\\ &\delta_{13}=2.3~(132^{\circ})\\
\end{aligned}$&$\begin{aligned}
	&m_4^2=3.507\text{eV}^2\\&\sin^2\theta_{14}=0.05,\\&\sin^2\theta_{24}=0.0067\\
	&\sin^2\theta_{34}=0.003\\
\end{aligned}$\\
\hline
		BMP-3&$\begin{aligned}
	&
	Y_1=1.087,~Y_2=1.294,\\&Y_3=0.043, Y_4=0.09\\&Y_5=0.48,Y_6=0.974\\
	&\langle \phi_1\rangle=246 \text{GeV}\\&\langle \phi_2\rangle\simeq0.01 \text{eV}\\
	&\kappa_1\sim\kappa_2\sim\kappa_3\simeq0.1,\\
	&q_1\sim q_2\simeq 0.002.
\end{aligned}$&$\begin{aligned}
	&
	\Delta m_{21}^2=6.98\times10^{-5}\text{eV}^2\\&\Delta m_{31}^2=2.47\times10^{-3}\text{eV}^2\\
	& \sin^2\theta_{12}=0.273\\ &\sin^2\theta_{13}=0.024,\\ &\sin^2\theta_{23}=0.460\\
	& \delta_{13}=5.01~(287^{\circ})\\
\end{aligned}$&$\begin{aligned}
	&m_4^2=1.337\text{eV}^2\\&\sin^2\theta_{14}=0.124,\\&\sin^2\theta_{24}=0.097\\
	&\sin^2\theta_{34}=0.0115\\
\end{aligned}$\\
\hline
	\end{tabular}
\caption{Three sets of benchmark points (BMPs) are shown here. With the chosen sets of model parameters, we display predictions of active and active-sterile neutrino mixing parameters. {  Here the $\delta_{13}$ values are showing in both radian and {\it degree} (in the parenthesis)}.}\label{prediction1}
\end{table}

{Combining these results from our model study, we can consider sterile mass around $0.1<m_4^2(\text{eV}^2)<4$ and mixing angles within the limits as, $1.7^{\circ}\le\theta_{14}\le13.29^{\circ}$, $0.35^{\circ}\le\theta_{24}\le10.9^{\circ}$ and $0.4^{\circ}\le\theta_{34}\le8.5^{\circ}$.} Current best fit value for $m_4^2$ is at 1.7 eV$^2$, and our results from the 5-zero texture in the $3+1$ scenario are in good agreement with it. Other mixing parameters results are shown in table \ref{com1} for better understanding. These results are yet not verified completely, however in future experiments soon we may get solid bounds on these parameters.  Along with the active-sterile mixing parameters, we keep table \ref{prediction1} for active and active-sterile benchmark points for the model parameters as input parameters. Without loss of generality, we keep fixed VEV for the scalar fields and fixed coupling value for $\kappa_i$ and $q_i$. Since Yukawa couplings ($Y_i$) play a significant role in our model study, we varied them within the allowed bounds (0.01-1) and evaluated neutrino parameters. We keep three sets of BMPs, which satisfy the latest bounds on neutrino parameters only.

\subsection{Baryogenesis via resonant leptogenesis}
\begin{figure}[h]
\includegraphics[scale=.4]{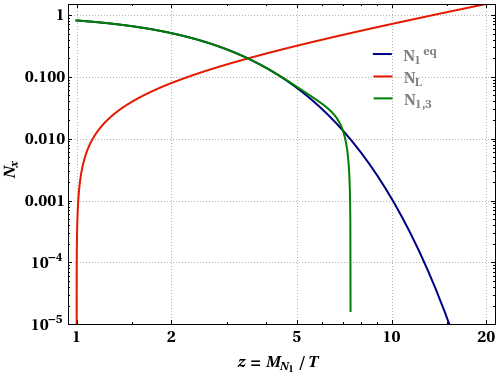}
\includegraphics[scale=.45]{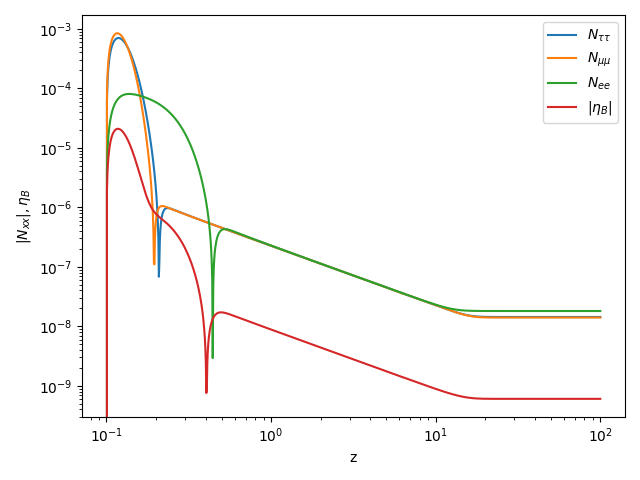}
\caption{(Left) Evolution of total lepton number density, RH neutrino density and its equilibrium number density with $z=M_{N_1}/T$. (Right) Evolution of lepton flavor asymmetries generated for three flavours and the current baryon asymmetry of the Universe. For RL, we have considered nearly degenerate mass spectrum for the heavy RH neutrinos $M_{N_1}\sim$ 1 TeV and a mass splitting equivalent to the decay width $\Delta M\sim \Gamma$.}\label{nd}
\end{figure}

In our numerical analysis, we will consider scenarios with two nearly degenerate heavy Majorana neutrinos $N_{1,3}$ with masses at the TeV range. The mass of the third heavy Majorana neutrino $N_2$ is taken to be of order $10^{7}$ GeV, so naturally, $N_2$ decouple from the low-energy sector of the theory. One flavor RL is not studied here, as it requires decaying RH mass $\mathcal{O}(10^{12})$ GeV \cite{DeSimone:2007edo}. { We analyze} the Boltzmann equations by numerically solving the values of the lepton asymmetry $N_L$ and the heavy-neutrino number densities $N_{N_i}$ as functions of the parameter $z=M_{N_1}/T$.
We set mass for $M_{N_1}\simeq M_{N_3}\simeq1$ TeV with $\frac{M_{N_3}}{M_{N_1}}-1=10^{-11}$ for our calculation. {  The choice of mass splitting order is quite trivial in the case of the resonant leptogenesis scenario. For a pair of heavy Majorana neutrinos, the necessary and sufficient condition for which the leptonic asymmetries of order unity can take place have to satisfy the following conditions:
\begin{equation}
	M_{N_i}-M_{N_j}\sim \Gamma_{N_{i,j}}/2 ~~~\text{and}~~~ \frac{
		|\text{
			Im}(Y^\dagger Y)^2_{ij}|}{(Y^\dagger Y)_{ii} (Y^\dagger Y)_{jj}}\sim 1,
\end{equation}  where, $\Gamma_{N_{i}}$ indicates the decay width of $N_i$ species.} After solving for the RH and lepton number densities, we have shown evolution pattern in fig.\ref{nd}. { We also used the leptogenesis equation solving tool {\rm ULYSSES}\footnote{ {\rm ULYSSES} is a leptogenesis equation solver that we have used to carry out our analysis and parameter scan for the whole leptogenesis calculation.}\cite{Granelli:2020pim} to carry out our calculations and later we have crosschecked the results in {\tt MATHEMATICA}}.
\begin{figure}
\includegraphics[scale=.4]{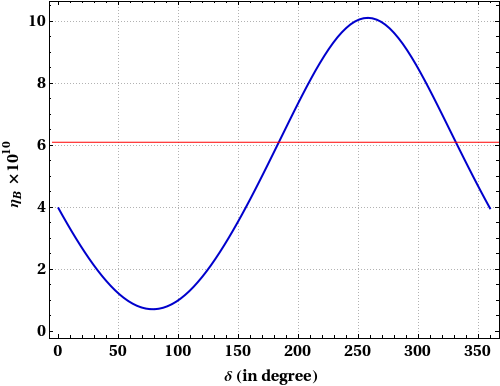}
\includegraphics[scale=.4]{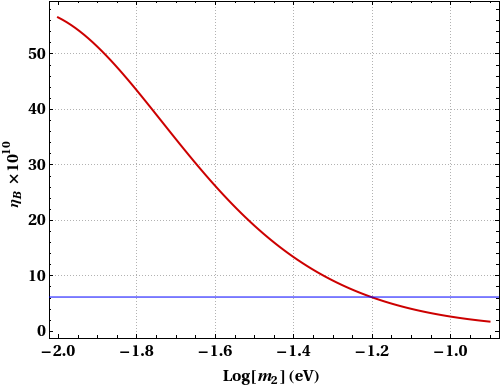}
\caption{In the left figure, variation for Dirac CP phase ($\delta$) with the observed BAU value is shown. $\delta$ around $180^{\circ}$ and $330^{\circ}$ does satisfy the current BAU value. On the right figure, we have varied lightest neutrino mass with $\eta_B$ value. Lightest neutrino mass satisfies the best fit $\eta_B$ value with $6.1\times 10^{-10}$ at 0.063 eV.}\label{del}
\end{figure}

Results show an obvious pattern for our model with the particle number densities. In the left panel of fig.\ref{nd}, evolution of RH neutrino density ($N_{1,3}$) and lepton density along with the equilibrium number density of $N_1$ are projected against $z$. In the right panel, variations of lepton number density for three flavours and $\eta_B$ with $z$ are shown. The initial abundance for the leptonic and baryon number densities was assumed to be zero, and over time by the decay of RH neutrinos, their concentration keeps increasing, at the same time the $N_1$ number density keeps decreasing. As soon as the out-of-equilibrium is achieved, the decay process slows down, and the inverse decay excels, which never comes into thermal equilibrium. Thus the green curve deviates from the blue curve, and the lepton asymmetry is generated with increasing number density.

For $M_{N_1}\sim$ TeV, we find that contribution from each lepton flavour is equally shared to give rise to the final lepton asymmetry. We have not shown the contributions from the off-diagonal terms ($N_{ij}$ with $i,j=e,\mu,\tau$ but $i\ne j$), as diagonal terms would dominate the whole situation here, due to the choice of our parameter space. Moreover, among the diagonal contributions, $N_{ee}$ curve kink hits the lowest value near $z\sim1$ is just a consequence of the values of the input parameters. However, there is a saturation region beyond $z=1$ for all flavour contributions and they are coherent due to the mass of the RH neutrino ($N_{1,3}$). This result slightly contradicts the RL$_{\tau}$ case, the significant contribution was coming from the $\tau$ lepton flavour, and other charge lepton contributions had less influence due to the larger washout rates \cite{Dev:2014laa}. Moreover, from previous studies, it is clear that within a strong washout regime, the final lepton or baryon asymmetry is independent of the initial concentration \cite{Dev:2014laa, Pilaftsis:2003gt, Blanchet:2011xq}. Even if we start with very large initial lepton asymmetry, the final asymmetry is achieved for $z\sim1$ within RL formalism by rapidly washing out the primordial asymmetry. 

We also varied other oscillation parameters associated with our model to the baryogenesis result. These model parameters are dictating the baryogenesis result through the CI parametrization used in equation \eqref{ci} via the $U_{PMNS}$ matrix. In fig. \ref{del}, a variation of the Dirac CP-phase with the observed BAU value is shown in the left panel. Delta ($\delta$) value around $180^{\circ}$ and $330^{\circ}$ are consistent with the current observed value, which is $\eta_B=(6.1\pm0.18)\times 10^{-10}$. In the right panel, we checked the $next$ lightest neutrino mass\footnote{Within MES the lightest neutrino mass is zero naturally, so the second lightest neutrino eigenvalue ($m_2$) is considered as lightest with a definite value.} bound with the baryogenesis result. For an increase in $m_2$, there is a gradual decrease in the $\eta_B$ value. With our choice of parameter space, the $next$ lightest neutrino mass is coming out as 0.063 eV with the baryogenesis bound.   
\subsection{Neutrinoless double beta decay ($0\nu\beta\beta$)}
\begin{figure}[h]\centering
\includegraphics[scale=0.49]{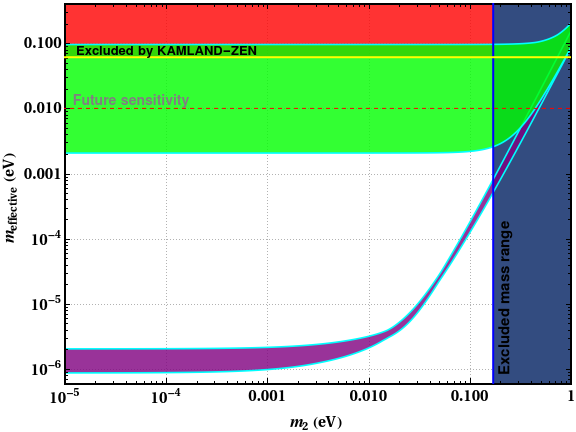}
\caption{Results of effective mass {\it vs.} the lightest neutrino mass. Green shaded region represents the effective mass contribution from active+sterile case and the purple region gives only active neutrino contribution.}\label{ndplot1}
\end{figure}
In this section, we study the numerical consequences of $0\nu\beta\beta$ using the bounds obtained from the previous section's results. We use global fit light neutrino parameters for active neutrinos from table \ref{tab:d1}, sterile parameters from texture zero bounds and CP phase from baryogenesis result. {  We have assumed zero Majorana phases ($\alpha=\beta=\gamma=0$) throughout our analysis.} Variation of effective mass with the lightest neutrino mass for active neutrinos and active+sterile neutrino contributions are shown in fig. \ref{ndplot1}. The red region above the horizontal yellow line represents the upper bound on effective mass given by KAMLAND-ZEN, and the dashed grey line gives the future sensitivity of the upper bound. The Blue region on the right side of the horizontal blue line gives the upper bound on the sum of all three active neutrinos ($=0.12$ eV). A much wider region (green) satisfying the effective mass is achieved in the case of active+sterile neutrino contribution, whereas a thin region (purple) is observed for active neutrino contribution only. Hence, a strong and impressive contribution from the sterile sector is observed in our 5-zero texture structure. Even though there is a wider range covered in the presence of sterile neutrino, it goes beyond the current upper bound by KAMLAND-ZEN. {  From previous studies \cite{Abada:2018qok,Das:2019kmn}, it was clear that a small mixing angle between active and sterile flavour can resolve the issue of violating the KAMLAND-ZEN upper bound on effective mass.}
\subsection{Dark matter}
\begin{figure}[h]
	\includegraphics[scale=.52]{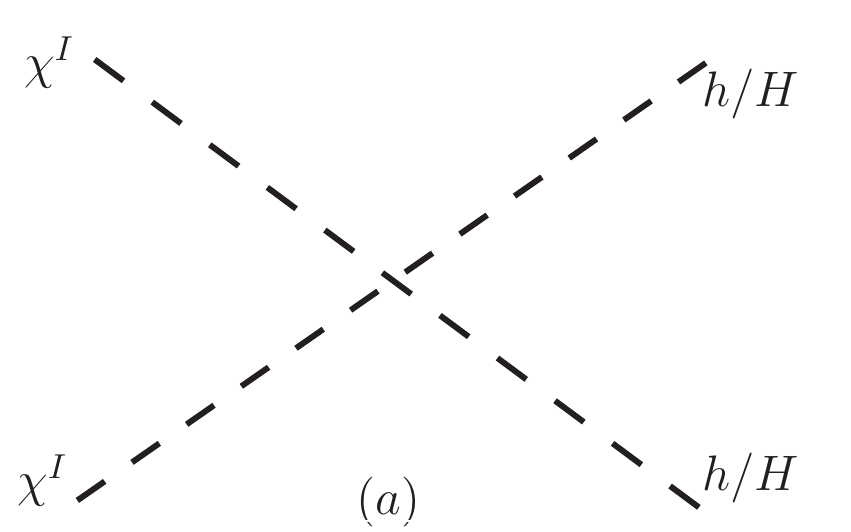}
	\includegraphics[scale=.52]{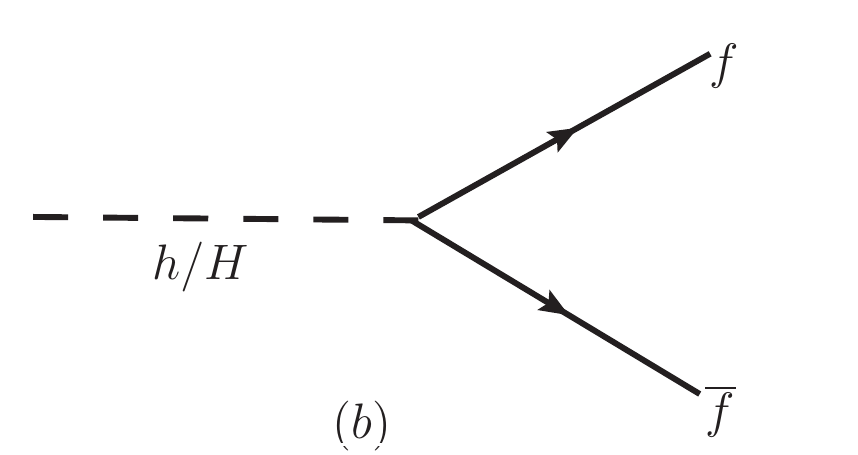}
	\includegraphics[scale=.52]{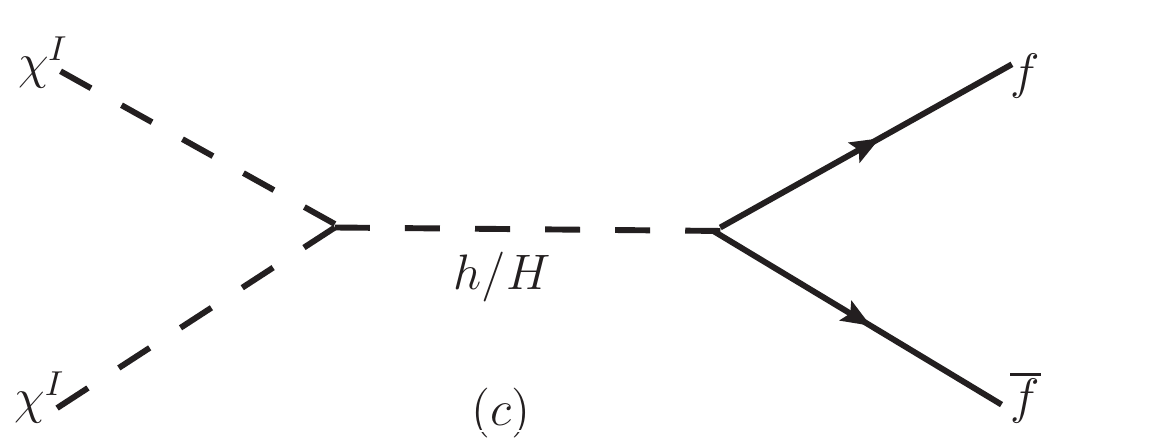}
	\caption{Annihilation processes of (a) $\chi^I\chi^I\rightarrow h/H$, $h/H\rightarrow SM~particles$ and Higgs mediated $\chi^I\chi^I\rightarrow SM~particles$. Here $f$ represents SM fermions.}\label{ann1}
\end{figure}

In this section, we will discuss various regions of DM parameter space, satisfying the current relic density. The relic density in this model mainly comes through the annihilation channels ($\chi^I\chi^I\rightarrow SM ~particle$), the intact $Z_{3,4}$ symmetry on $\chi^I$ and the mass gap ($M_{\chi^I}-M_H, M_{\chi^I}-M_{N_1}$) do not allow the co-annihilation diagrams. Also, the decay of the $\chi^I$ is greater than the lifetime of the Universe ($t_{\chi^I}\gg t_{Universe}$). The annihilation diagrams are shown in fig. \ref{ann1}.
\begin{figure}[h]
	\includegraphics[scale=0.29]{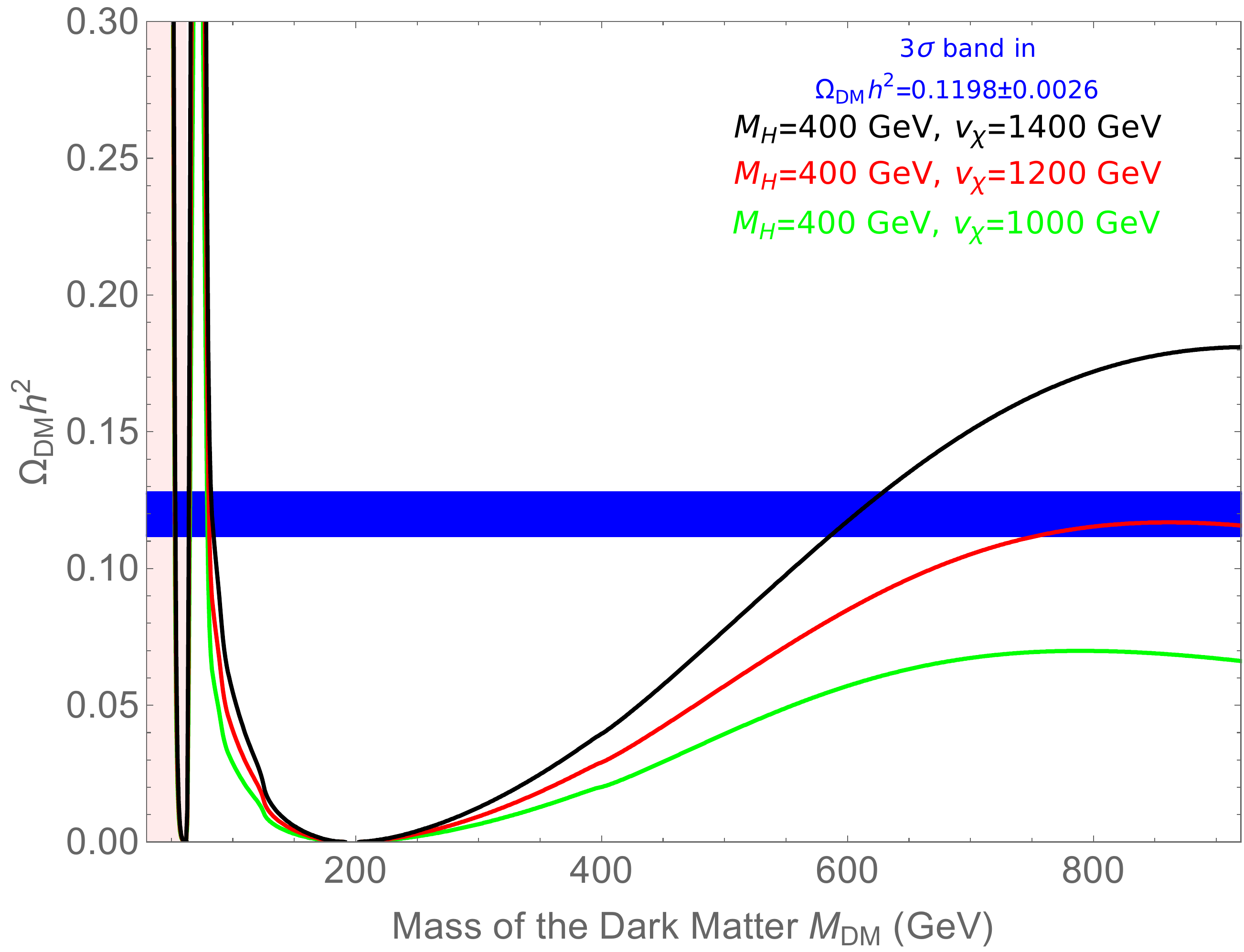}
	\includegraphics[scale=0.29]{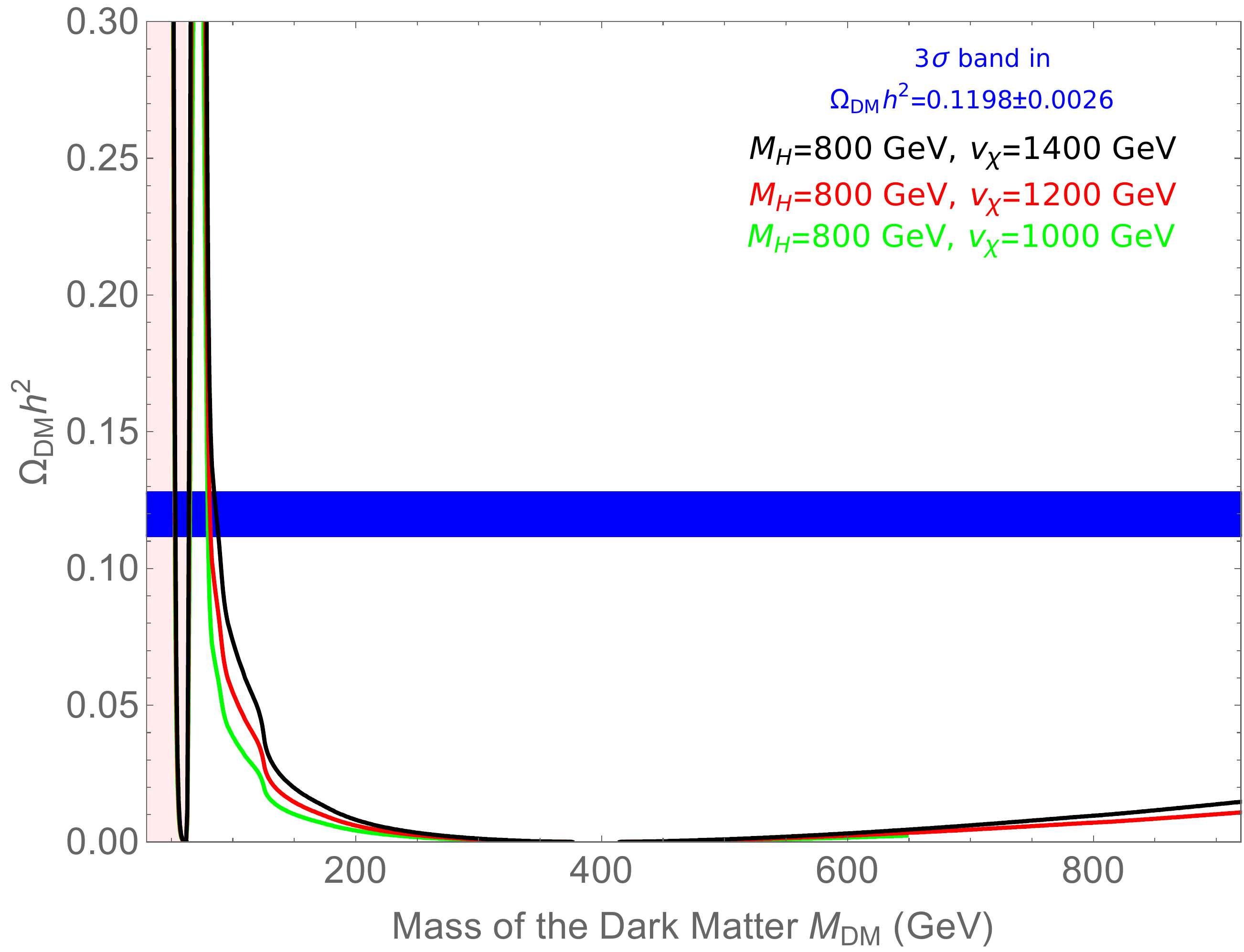}
	\caption{Dark matter density against dark matter mass for two different heavy Higgs masses ($M_H=400$ GeV and 800 GeV). We took three different VEV of the singlet scalar and the corresponding variation is shown as green (1000 GeV), red (1200 GeV) and black (1400 GeV). The blue band stands for 3
		4$\sigma$ variation of experimental relic density data.}\label{dm1}
\end{figure}
\begin{table}[h]
	\begin{tabular}{|c|c|c|c|c|c|p{4cm}|}
		\hline
		BMP&$M_{DM}$ (GeV)&$M_H$ (GeV)&$v_{\chi}$ (GeV)&$\cos\alpha$&$\Omega_{DM}h^2$& Processes\\
		\hline
		I&65.4&400&1000&0.95&0.119&$\chi^I\chi^I\rightarrow b\bar{b}$(67\%) 
		$\chi^I\chi^I\rightarrow W^+W^-(22\%)$\\
		\hline
		II&200&208&1000&0.95&0.118&$\chi^I\chi^I\rightarrow W^+W^-$(40\%)
		$\chi^I\chi^I\rightarrow HH~(32\%)$ $\chi^I\chi^I\rightarrow ZZ~(19\%)$\\
		\hline
		III&800&200&1000&0.95&0.121&$\chi^I\chi^I\rightarrow SS(79\%)$
		$\chi^I\chi^I\rightarrow HS(16\%)$\\
		\hline	\end{tabular}
	\caption{Benchmark points for various processes at different dark matter mass and $S$ fermion mass with relic density for $v_{\chi}=1000$ GeV.}\label{bmpt1}
\end{table}

\begin{table}[h]
	\begin{tabular}{|c|c|c|c|c|c|p{4cm}|}
		\hline
		BMP&$M_{DM}$ (GeV)&$M_H$ (GeV)&$v_{\chi}$ (GeV)&$\cos\alpha$&$\Omega_{DM}h^2$& Processes\\
		\hline
		I&53.06&400&1200&0.95&0.121&$\chi^I\chi^I\rightarrow b\bar{b}$(77\%) 
		$\chi^I\chi^I\rightarrow W^+W^-(12\%)$\\
		\hline
		II&200&218&1200&0.95&0.121&$\chi^I\chi^I\rightarrow W^+W^-$(41\%)
		$\chi^I\chi^I\rightarrow HH~(30\%)$ $\chi^I\chi^I\rightarrow ZZ~(19\%)$\\
		\hline
		III&800&397&1200&0.95&0.118&$\chi^I\chi^I\rightarrow SS(30\%)$
		$\chi^I\chi^I\rightarrow W^+W^-(29\%)$
		$\chi^I\chi^I\rightarrow HH(21\%)$
		$\chi^I\chi^I\rightarrow ZZ(14\%)$\\
		\hline	\end{tabular}
	\caption{Benchmark points for various processes at different dark matter mass and $S$ fermion mass with relic density for $v_{\chi}=1200$ GeV.}\label{bmpt2}
\end{table}

\begin{table}[h]
	\begin{tabular}{|c|c|c|c|c|c|p{4cm}|}
		\hline
		BMP&$M_{DM}$ (GeV)&$M_H$ (GeV)&$v_{\chi}$ (GeV)&$\cos\alpha$&$\Omega_{DM}h^2$& Processes\\
		\hline
		I&53.2&380&1400&0.95&0.122&$\chi^I\chi^I\rightarrow b\bar{b}$(77\%) 
		$\chi^I\chi^I\rightarrow W^+W^-(12\%)$\\
		\hline
		II&200&228&1400&0.95&0.118&$\chi^I\chi^I\rightarrow W^+W^-$(41\%)
		$\chi^I\chi^I\rightarrow HH~(29\%)$ $\chi^I\chi^I\rightarrow ZZ~(19\%)$\\
		\hline
		III&800&450&1400&0.95&0.118&$\chi^I\chi^I\rightarrow W^+W^-(39\%)$
		$\chi^I\chi^I\rightarrow HH(24\%)$
		$\chi^I\chi^I\rightarrow ZZ(19\%)$
		$\chi^I\chi^I\rightarrow SS(14\%)$\\
		\hline	\end{tabular}
	\caption{Benchmark points for various processes at different dark matter mass and $S$ fermion mass with relic density for $v_{\chi}=1400$ GeV.}\label{bmpt3}
\end{table}

\begin{table}[h]
	\begin{tabular}{|c|c|c|c|c|c|p{4cm}|}
		\hline
		BMP&$M_{DM}$ (GeV)&$M_H$ (GeV)&$v_{\chi}$ (GeV)&$\cos\alpha$&$\Omega_{DM}h^2$& Processes\\
		\hline
		I&54.7&800&3000&0.95&0.119&$\chi^I\chi^I\rightarrow b\bar{b}$(76\%) 
		$\chi^I\chi^I\rightarrow W^+W^-(13\%)$\\
		\hline
		II&134&800&3000&0.95&0.120&$\chi^I\chi^I\rightarrow W^+W^-$(50\%)
		$\chi^I\chi^I\rightarrow HH~(29\%)$ $\chi^I\chi^I\rightarrow ZZ~(21\%)$\\
		\hline
		III&940&800&4000&0.95&0.119&$\chi^I\chi^I\rightarrow W^+W^-(48\%)$
		$\chi^I\chi^I\rightarrow ZZ(24\%)$
		$\chi^I\chi^I\rightarrow HH(23\%)$\\
		\hline	\end{tabular}
	\caption{Benchmark points for various processes at different dark matter mass and $S$ fermion mass with relic density for $v_{\chi}>3000$ GeV.}\label{bmpt4}
\end{table}
The coupling strength for the interaction $\chi^I\chi^Ihh(HH)$ is $\lambda_2\cos\alpha\sin\alpha$ and the Higgs portal couplings are $g_{\chi^I\chi^Ih}=\frac{\sin\alpha(M^2_{DM}+M_h^2)}{v_{\chi}}$ and $g_{\chi^I\chi^IH}=-\frac{\cos\alpha(M_{DM}^2+M_H^2)}{v_{\chi}}$. It is clear that the dark matter relic density and direct detection cross-section mainly depend on $M_{DM}$, VEVs ($v,v_{\chi}$), mixing angle $\alpha$ and other quartic couplings $\lambda$. Hence, these are depending on the masses of the scalar particles too. In this analysis we will keep the mixing angle fixed at $\cos\alpha=0.95$ and vary other parameters, mainly, $v_{\chi},M_H$ and $m_{DM}$. The $\cos\alpha=0.95$ ensures the Higgs signal strength within the experimental limits \cite{Bechtle:2013xfa,Sirunyan:2018ouh}. In the fig \ref{dm1}, we have varied the DM mass, $M_{DM}$ along x-axis and plotted the corresponding relic density in the y-axis. We keep fixed $M_H$ at 400 GeV and 800 GeV in the respective figures. For three different values of singlet scalar VEVs $v_{\chi}=1000,1200,1400$ GeV, we have carried out the whole DM analysis. The blue band indicates the relic density band at current 3$\sigma$, $\Omega_{DM}h^2=0.1198\pm0.0026$ \cite{Aghanim:2018eyx}. One can see from these figures that the relic density in this model can be obtained near $M_{DM}\sim M_h/2$ region. This is the Higgs ($h$) resonance region, we need vary small Higgs portal coupling to get the relic density. The other coupling produces a very large $\langle\sigma v\rangle$, hence, a depletion region is occurred near $M_{DM}\sim M_h/2$. In this region $\chi^I\chi^I\rightarrow h b\bar{b}$ is the dominating annihilation process. Near $|M_{DM}-M_h|<3$ GeV region, the cross section $\langle\sigma v\rangle$ remains small due to the small $h$-portal coupling, hence we get very large relic density ($\Omega_{DM}h^2\propto\frac{1}{\langle \sigma v\rangle}$). It is to be noted that the Higgs ($H$) portal coupling, {\it i.e.}, $H$-mediated diagram has a tiny effect in this region. After $M_{DM}\simeq70$ GeV, $\chi^I\chi^I\rightarrow h\rightarrow VV^*~(V=W,Z)$ starts dominates over other annihilation channels ($V^*$ stands for virtual gauge boson). $\chi^I\chi^I\rightarrow h\rightarrow WW$ become effective at $M_{DM}>M_W$, whereas $\chi^I\chi^I\rightarrow h\rightarrow ZZ$ become effective at $M_{DM}>M_Z$. 

The effective annihilation cross-section becomes large at $M_{DM}>70$ GeV; hence the relic density becomes small. One can understand these effects from fig \ref{dm1}. The relic density $\Omega_{DM}h^2$ falls again near $M_{DM}\sim M_H/2$ where $\sigma(\chi^I\chi^I\rightarrow H\rightarrow XX(X=W,Z,h))$ cross-section become large in this $H$-resonance region. The relic density is again starting to increase after $M_H/2$ with the mass of DM as $\Omega_{DM}h^2\propto M_{DM}$ too. In the left figure of \ref{dm1}, we consider small $H$-scalar mass $M_H=400$ GeV, hence the Higgs portal coupling remains small as compared to $M_H=800$ GeV. Also the larger $v_{\chi}=1200$ and 1400 GeV gives suppression to provide exact relic density for $M_H=400$ GeV. However, we were unable to achieve relic for $M_H=800$ GeV in the high mass region. It is to be noted that we can get the exact relic density for $M_H=800$ GeV for very large $v_{\chi}>3000$ GeV. We have presented various benchmark points allowing the current relic density value and present direct detection data with corresponding contributions in tables \ref{bmpt1}-\ref{bmpt4} (DM annihilation processes with contributions below 10\% are not shown in the chart.).

\section{Conclusion}\label{sec5}
In this work, we explore the possibility of five zero textures in the active-sterile mixing matrix under the framework of the minimal extended seesaw (MES). In the $4\times4$ mixing matrix, elements in the fourth column are restricted to be zero; hence, we impose zeros only in the active neutrino block. There are six possible structures with five zeros, and among them, only one is allowed ($T_5$) by the current oscillation data in the normal hierarchy mass ordering. There is a broken $\mu-\tau$ symmetry as $M^{\nu}_{\mu4}\ne M^{\nu}_{\tau4}$; hence, a non-zero reactor mixing angle ($\theta_{13}$) can also be achieved from this structure, which we skip in our study. We constructed the desirable mass matrices with the help of discrete flavour symmetries like $A_4,Z_3$ and $Z_4$. In addition to the SM particle content three RH neutrinos, a single fermion and an additional low-scaled Higgs doublet are considered. This additional Higgs gets VEV via soft breaking mass term. A $A_4$ triplet flavons $\psi$ is associated with the diagonal charged lepton and Dirac mass matrix generation. Singlet flavons $\eta_1,\eta_2$ and $\chi$ were related to the Majorana and sterile mass generation, respectively. The flavon $\chi$ is considered to be a complex singlet, and the imaginary component of this ($\chi^I$) will behave as a dark matter candidate in this study.

The active neutrino part is skipped in this study, and we mainly focused on active-sterile bounds from the five zero textures. Notable bounds on the sterile mass and the active-sterile mixing angles are obtained in this study. In comparison with the global 3$\sigma$ results, we get sterile mass between $0.1<m_4^2 (\text{eV}^2)<4$ and other mixing angles, as shown in table \ref{com1}, which agrees with current experimental results. We expected to see verification/falsification of these model bounds from texture-zeros in future experiments.

A nearly degenerate mass pattern for the RH neutrinos is considered in such a way that they can exhibit resonant leptogenesis at the TeV scale. For successful RL leptogenesis, the mass splitting of the RH neutrinos should be of the order of the decay width of the particle, and we choose $\frac{M_{N_3}}{M_{N_1}}-1\sim10^{-11}$ in our study. {  Semi-classical approach to the Boltzmann equation is used and solved to see the evolution of the particles in out-of-equilibrium conditions.} We have included flavour effects in this study and with our choice of mass, we get an equal contribution from each charged lepton flavour in the final lepton asymmetry due to the choice of our input mass of the RH neutrinos ($N_{1,3}\sim1$ TeV). We also check bound from baryogenesis on the Dirac CP phase ($\delta$) as well as the lightest neutrino mass and we get $\delta$ around $180^{\circ}$ and $330^{\circ}$ while $m_{lightest}$ on 0.069 eV with current best fit value $\eta_B=6.1\times10^{-10}$.  

With the bounds obtained from texture zero, baryogenesis study and available data from global fit results for light neutrino parameters, we study the influence of sterile neutrino in the effective mass calculation. A large enhanced region from sterile contribution is obtained in comparison to the SM contribution. The sterile contribution goes beyond the current upper bound of the effective mass obtained from various experiments, which gives support in favour of minimal active-sterile mixing angle in future works.

We studied the allowed parameter space of the model, taking into account various theoretical bounds for dark matter mass $10$ GeV to 1000 GeV. The imaginary part of the complex singlet $\chi$ serves as a potential dark matter candidate in this study. Due to the choice of the mass spectrum of the particles, only the annihilation channels are contributing to the relic density through Higgs portal coupling ($h/H$). The Higgs portal mixing angle $\alpha$, $M_H$ and the VEV of $\chi$ ($v_{\chi}$) and the quartic couplings play a significant role in dark matter analysis. We have chosen three sets of $v_{\chi}$'s and two sets of $M_H$s specifically to visualize various dark matter mass regions satisfying current relic abundance. In the lower dark matter mass region ($\sim M_h/2$), both the cases satisfy relic abundance due to the Higgs resonance. However, in the High mass regions, only in the $M_H=400$ GeV case, we do get a satisfying relic due to the small Higgs suppression. In the $M_H=800$ GeV case, a very large VEV of $\chi$ does satisfy the current relic abundance value. 

In the final word, neutrino mass generation does not have a direct connection to the dark matter and matter asymmetry studies under the tree-level seesaw framework. However, with flavour symmetry, the complex singlet scalar $\chi$ and the RH neutrinos are the bridges to connect these sectors under the same roof. We can see how the VEV $v_{\chi}$ controls the dark matter mass throughout various regimes as an immediate consequence of flavour symmetries. 
\section{Acknowledgement} PD and MKD would like to acknowledge the Department of Science and Technology,  Government of India under project number EMR/2017/001436 for financial aid. NK would like to thank Dilip Kumar Ghosh for his support at IACS, Kolkata.
\appendix
\section{The Boltzmann equation}\label{appx}
The BEs in \eqref{be1} includes the collision processes like $N_j\rightarrow L_i\phi$ as well as $L_k\phi\leftrightarrow L_i\phi$ and $L_k\phi\leftrightarrow L_i^c\phi^{\dagger}$ scattering processes, which are defined as \cite{Pilaftsis:2003gt}
\begin{eqnarray}
	\gamma_{L\phi}^{N_j}&\equiv& \sum_{k=e,\mu,\tau}\Big[\gamma(N_j\rightarrow L_k\phi)+\gamma(N_j\rightarrow L_k^c\phi^{\dagger})\Big],\\
	\gamma_{L_i\phi}^{L_k\phi}&\equiv&\gamma(L_k\phi\rightarrow L_i\phi)+\gamma(L_k^c\phi^{\dagger}\rightarrow L^c_i\phi^{\dagger}),\\
	\gamma_{L_i^c\phi^{\dagger}}^{L_k\phi}&\equiv&\gamma(L_k\phi\rightarrow L_i^c\phi^{\dagger})+\gamma(L_k^c\phi^{\dagger}\rightarrow L_i\phi).
\end{eqnarray}
Including the contributions from narrow width approximation (NWA) as well as real intermediate states (RISs) from \cite{Pilaftsis:2003gt}, these collision terms are explained as,
\begin{eqnarray}
	&\gamma_{L\phi}^{N_j}=&\frac{m_N^3}{\pi^2z}K_1(z)\Gamma_{N_j},\\
	&\gamma_{L_i\phi}^{L_k\phi}=&\sum_{\alpha,\beta}^{3}(\gamma_{L\phi}^{N_{\alpha}}+\gamma_{L\phi}^{N_{\beta}})\frac{2\big(\bar{Y^*_{i\alpha}}\bar{Y_{k\alpha}^{c*}}\bar{Y_{i\beta}}\bar{Y^{c*}_{k\beta}}+\bar{Y^{c*}_{i\alpha}}\bar{Y_{k\alpha}^{*}}\bar{Y_{i\beta}}\bar{Y^{c}_{k\beta}}\big)}{\Big[(\bar{Y^{\dagger}}\bar{Y})_{\alpha\alpha}+(\bar{Y^{c\dagger}}\bar{Y^c})_{\alpha\alpha}+(\bar{Y^{\dagger}}\bar{Y})_{\beta\beta}+(\bar{Y^{c\dagger}}\bar{Y^c})_{\beta\beta}\Big]^2}\nn\\
	&&\times\Big(1-2i\frac{M_{N_{\alpha}}-M_{N_{\beta}}}{\Gamma_{N_{\alpha}}+\Gamma_{N_{\beta}}}\Big)^{-1},\\
	&\gamma_{L_i^{c}\phi^{\dagger}}^{L_k\phi}=&\sum_{\alpha,\beta=1}^{3}(\gamma_{L\phi}^{N_{\alpha}}+\gamma_{L\phi}^{N_{\beta}})\frac{2\big(\bar{Y^*_{i\alpha}}\bar{Y_{k\alpha}^{*}}\bar{Y_{i\beta}}\bar{Y_{k\beta}}+\bar{Y^{c*}_{i\alpha}}\bar{Y_{k\alpha}^{*}}\bar{Y_{i\beta}^c}\bar{Y^{c}_{k\beta}}\big)}{\Big[(\bar{Y^{\dagger}}\bar{Y})_{\alpha\alpha}+(\bar{Y^{c\dagger}}\bar{Y^c})_{\alpha\alpha}+(\bar{Y^{\dagger}}\bar{Y})_{\beta\beta}+(\bar{Y^{c\dagger}}\bar{Y^c})_{\beta\beta}\Big]^2}\nn\\
	&&\times\Big(1-2i\frac{M_{N_{\alpha}}-M_{N_{\beta}}}{\Gamma_{N_{\alpha}}+\Gamma_{N_{\beta}}}\Big)^{-1}.
\end{eqnarray}
Finally the BEs of \eqref{be1} are rewrite in the form,
\begin{eqnarray}
	\nn&\frac{d\eta_{L_i}}{dz}=&\frac{z}{\eta_{\gamma}H(z=1)}\Big[\sum_{j=1}^{3}\big(\frac{\eta_{N_j}}{\eta_N^{eq}}-1\big)\epsilon_i\gamma_{L\phi}^{N_j}\\
	\nn&&-\frac{2}{3}\eta_{L_i}\Big\{\sum_{j=1}^{3}\gamma_{L\phi}^{N_j}B_{ij}+\sum_{k=e,\mu,\tau}\big(\gamma_{L_i^{c}\phi^{\dagger}}^{\prime L_i\phi}+\gamma_{L_k\phi}^{\prime L_i\phi}\big)\Big\}\\
	&&-\frac{2}{3}\sum_{j=1}^{3}\Big\{\eta_{L_k}\epsilon_{ii}\epsilon_{ik}\gamma_{L\phi}^{N_j}B_{ij}+\big(\gamma_{L_i^{c}\phi^{\dagger}}^{\prime L_i\phi}-\gamma_{L_k\phi}^{\prime L_i\phi}\big)\Big\}\Big].\label{be5}
\end{eqnarray}

Here, $\gamma_Y^{\prime X} =\gamma_Y^X-(\gamma_Y^X)_{\text{RIS}}$ denote the RIS=subtraction collision terms which are motivated from past studies \cite{Pilaftsis:2003gt, Deppisch:2010fr} and $B_{ij}$ are the branching rations,
\begin{eqnarray}
	B_{ij}=\frac{|\bar{Y}_{ij}|^2+|\bar{Y}_{ij}^c|^2}{(\bar{Y^{\dagger}\bar{Y}})_{ii}+(\bar{Y}^{c\dagger}\bar{Y^c})_{ii}}.
\end{eqnarray} 

\section{The full scalar potential}\label{apotential}
Six scalar flavons are in our model. The complete structure of the potential would be,
\begin{eqnarray}
	V=\nn&&-\mu_{\phi_1}^2 \, (\phi_1^{\dagger}\phi_1)-\mu_{\phi_2}^2\, (\phi_2^{\dagger}\phi_2)+\mu_{\phi_1\phi_2}^2(\phi_1^{\dagger}\phi_2+h.c.)+\lambda_{1h}(\phi_1^{\dagger}\phi_1)^2+\lambda_{2h} (\phi_2^{\dagger}\phi_2)^2\\\nn
	&&+\lambda_{3h}(\phi_1^{\dagger}\phi_1)(\phi_2^{\dagger}\phi_2)+\lambda_{4h}(\phi_1^{\dagger}\phi_2)(\phi_2^{\dagger}\phi_1)-\mu_{\psi_1}^2 (\psi_1^{\dagger}\psi_1)-\mu_{\psi_2}^2 (\psi_2^{\dagger}\psi_2) -\mu_{\eta}^2 (\eta^{\dagger}\eta)\\\nn
	&&+\frac{\lambda_{\psi_1}}{4!}(\psi_1^{\dagger}\psi_1)^2+\frac{\lambda_{\eta}}{4!}(\eta^{\dagger}\eta)^2+\lambda_{\psi_1\phi_1}(\phi_1^{\dagger}\phi_1)(\psi_1^{\dagger}\psi_1)+\lambda_{\eta\phi_1}(\phi_1^{\dagger}\phi_1)(\eta^{\dagger}\eta)\\\nn
	&& {+\mu^2_{\psi_{1,2}}(\psi_1^\dagger\psi_2)+\lambda_{\psi_{1,2}}(\psi_1^\dagger\psi_2)^2+\lambda_{\psi_1\eta}(\psi_1^\dagger\psi_1)(\eta^\dagger\eta)}\\\nn
	&&+\lambda_{\psi_1\phi_2}(\phi_2^{\dagger}\phi_2)(\psi_1^{\dagger}\psi_1)+\lambda_{\eta\phi_2}(\phi_2^{\dagger}\phi_2)(\eta^{\dagger}\eta) -\frac{1}{2} \mu_{\chi^R}^2 \chi^{\dagger} \chi - \frac{1}{2}  \mu_{\chi^I}^2 (\chi^2 + h.c) \\
	&&+\frac{\lambda_{\chi}}{4!}(\chi^{\dagger}\chi)^2+\frac{\lambda_{\psi_2}}{4!}(\psi_2^{\dagger}\psi_2)^2+\lambda_{\chi\phi_1}(\phi_1^{\dagger}\phi_1)(\chi^{\dagger}\chi)+\lambda_{\chi\phi_2}(\phi_2^{\dagger}\phi_2)(\chi^{\dagger}\chi)\\\nn
	&&+\lambda_{\psi_2\phi_1}(\phi_1^{\dagger}\phi_1)(\psi_2^{\dagger}\psi_2)+\lambda_{\psi_2\phi_2}(\phi_2^{\dagger}\phi_2)(\psi_2^{\dagger}\psi_2)+\lambda_{\chi\psi_1}(\chi^{\dagger}\chi)(\psi_1^{\dagger}\psi_1)+\lambda_{\psi_1\psi_2}(\psi_2^{\dagger}\psi_2)(\psi_1^{\dagger}\psi_1)\\\nn
	&&+\lambda_{\chi\eta}(\chi^{\dagger}\chi)(\eta^{\dagger}\eta)+\lambda_{\eta\psi_2}(\psi_2^{\dagger}\psi_2)(\eta^{\dagger}\eta)+\lambda_{\chi\psi_2}\{(\psi_2^{\dagger}\psi_2)(\chi^{\dagger}\chi)\}\nn
\end{eqnarray}

We already discuss about the transformation of { these scalars} under $A_4$ symmetry.  The $A_4$ symmetry breaks when the fields get VEVs.  {We would like to mention that the softly $Z_4$ breaking term $\mu^2_{\phi_1 \phi_2} (\phi_1^\dagger \phi_2+ h.c)$ term will help to get non-zero pseudoscalar masses. And the part $\chi^2+h.c=(\chi^R)^2-(\chi^I)^2 + 2 \chi^R v_{\chi} + v_{\chi}^2$ will help to get the no-zero mass for the  singlet type pseudoscalar.} After symmetry breaking, the scalar field $\phi_1$ gets VEV $v_h$ while $\phi_2$ gets a tiny VEV ($v_{2h}\sim0.1$ eV) via soft-breaking term~\cite{Davidson:2009ha}. It satisfies $\sqrt{v_h^2+v_{2h}^2}\approx 246$ GeV.
These scalar fields after EWSB can be expressed as,
\begin{eqnarray}
	\phi_1&=&\frac{1}{\sqrt{2}}\begin{pmatrix}
		\phi_h^+ \\ \phi_{h} + v_h+i \phi_h^0
	\end{pmatrix},
	\, \, \phi_2=\frac{1}{\sqrt{2}}\begin{pmatrix}
		\phi_{2h}^+\\ \phi_{2h} + v_{2h}+i \phi_{2h}^0
	\end{pmatrix}\\
	\chi &=&(\chi^R+v_{\chi}+i\chi^I)/\sqrt{2},\,\, \eta= (\eta^R+u)/\sqrt{2},\,\, \psi_i= (\psi_i^R+v)/\sqrt{2}
\end{eqnarray}
In our study, the scalar potential related to these fields decouples from each other and can be understood as follows.
Let us calculate the details of the scalar potential.
The minimization conditions are given by 
\begin{eqnarray}
	\mu _{\phi _1}^2 &=& \frac{1}{2 v_h} \left\{ 2\lambda_{h}v_h^3+v_h \left((\lambda _{3 h}+\lambda _{4 h}) v_{2 h}^2+ \lambda _{\eta \phi _1}u^2+\lambda _{\chi \phi _1}v_{\chi}^2 + \lambda _{\psi _1 \phi _1}v^2+\lambda _{ \phi _1\psi_2} v^2\right)+2 \, v_{2 h} \, \mu _{\phi _1 \phi _2}^2 \right\},\nn\\
	\mu _{\phi _2}^2 &=& \frac{1}{2 v_h} \left\{ 2\lambda_{2h}v_h^3+ \left((\lambda _{3 h}+\lambda _{4 h}) v_{ h}^2v_{2h}+ \lambda _{\eta \phi _2}u^2+\lambda _{\chi \phi _1}v_{\chi}^2 + \lambda _{\psi _1 \phi _2}v^2+\lambda _{ \phi _2\psi_2} v^2\right)+2 \, v_{2 h} \, \mu _{\phi _1 \phi _2}^2 \right\},\nn\\
	\mu _{\chi ^R}^2 &=& \lambda _{\chi \phi _1}v_h^2 + \lambda _{\chi \phi _1}v_{2 h}^2-2 \mu _{\chi ^I}^2+\lambda _{\chi \eta }u^2+ \lambda _{\chi \psi_1}v^2+ \lambda _{\chi \psi _2}v^2+\frac{1}{12} \lambda _{\chi } v_{\chi }^2,\nn\\
	\mu _{\eta }^2 &=&  \lambda _{\eta \phi _1}v_h^2+ \lambda _{\eta\phi _2}v_{2 h}^2+\lambda _{ \eta\psi_2}v^2+\frac{1}{12} \lambda _{\eta}u^2+ \lambda _{\eta \psi _1}v^2+\lambda _{\chi \eta} v_{\chi }^2  +4  \lambda _{ \eta \psi _1 \psi _{2} } v^2,\\
	\mu _{\psi_1 }^2 &=& \lambda _{\psi_1 \phi _1}v_h^2+ \lambda _{\psi_2 \phi _1}v_{2h}^2+\lambda _{\eta\psi _1}u^2+\frac{1}{12} \lambda _{\psi_1}v^2+\lambda _{\psi_1 \psi_2}v^2+\lambda _{\chi \psi_1} v_{\chi }^2 + 2 \mu _{\psi _1 \psi _2}^2 +2  \lambda _{\eta \psi _1 \psi _{2}  }u^2 ,\nn\\
	\mu _{\psi_2}^2 &=& \lambda _{\psi_2 \phi _1}v_h^2+ \lambda _{\psi_2 \phi _2}v_{2h}^2+\lambda _{\eta\psi _2}u^2+\frac{1}{12} \lambda _{\psi_2}v^2+\lambda _{\psi_1 \psi_2}v^2+\lambda _{\chi \psi_2} v_{\chi }^2+2 \mu _{\psi _1 \psi _2}^2 +2  \lambda _{\eta \psi _1 \psi _{2}  } u^2 ,\nn\\
\end{eqnarray}
The CP-even scalar mass matrix ($\phi_h, \phi_{2h}, \chi^R, \psi_1, \psi_2$ and $\eta$) can be written as,
\begin{eqnarray}
	\mathcal{M}^2_{CP-even}&= { \fontsize{7}{10} \left(
		\begin{array}{cccccc}
			\mathcal{M}_{11}&\mathcal{M}_{12}& \lambda _{\chi \phi _1} v_h v_{\chi } &\lambda _{\phi _1 \psi _1} v v_h  &  \lambda _{\phi _1 \psi _2}v v_h & \lambda _{\eta \phi _1}u v_h  \\
			\mathcal{M}_{12} & \mathcal{M}_{22} &  \lambda _{\chi \phi _1}v_{2 h} v_{\chi } &  \lambda _{\phi _2 \psi _1}v v_{2 h} & \lambda _{\phi _2 \psi _2}  v v_{2 h}&  \lambda _{\eta \phi _2}u v_{2 h} \\
			\lambda _{\chi \phi _1} v_h v_{\chi } &  \lambda _{\chi \phi _1}v_{2 h} v_{\chi } & \frac{1}{12} \lambda _{\chi } v_{\chi }^2 & \lambda _{\chi \psi _1}v v_{\chi }  &  \lambda _{\chi \psi _2}v v_{\chi } & \lambda _{\chi \eta } u v_{\chi } \\
			\lambda _{\phi _1 \psi _1} v v_h &  \lambda _{\phi _2 \psi _1}v v_{2 h} &  \lambda _{\chi \psi _1}v v_{\chi } & \mathcal{M}_{44}& \mathcal{M}_{45}& \mathcal{M}_{46}\\
			\lambda _{\phi _1 \psi _2} v v_h  &  \lambda _{\phi _2 \psi _2}v v_{2 h} &  \lambda _{\chi \psi _2}v v_{\chi } & \mathcal{M}_{54}& \mathcal{M}_{55}&\mathcal{M}_{56}\\
			\lambda _{\eta \phi _1} u v_h &  \lambda _{\eta \phi _2}u v_{2 h} & \lambda _{\chi \eta } u v_{\chi } & \left(\lambda _{\eta \psi _1}+2 \lambda _{\eta  \psi _1 \psi _2}\right) u v & \left(\lambda _{\eta \psi _2}+2 \lambda _{\eta  \psi _1 \psi _2}\right) u v & \frac{ \lambda _{\eta }u^2}{12} \\
		\end{array}
		\right)}
	\label{eq:cpevenmass}
\end{eqnarray}
Where, $\mathcal{M}_{11}=  2  \lambda _hv_h^2-\frac{v_{2 h} \mu _{\phi _1 \phi _2}^2}{v_h} $, $\mathcal{M}_{12}= \mu _{\phi _1 \phi _2}^2+\left(\lambda _{3 h}+\lambda _{4 h}\right)v_h v_{2 h}  $, $\mathcal{M}_{22}=2  \lambda _{2 h}v_{2 h}^2-\frac{v_h \mu _{\phi _1 \phi _2}^2}{v_{2 h}}$, $\mathcal{M}_{44}=-\lambda _{\eta  \psi _1 \psi _2} u^2-\mu _{\psi _1 \psi _2}^2+\frac{1}{12}  \lambda _{\psi _1}v^2 $, $\mathcal{M}_{45}=\lambda _{\eta  \psi _1 \psi _2} u^2+\mu _{\psi _1 \psi _2}^2+ \lambda _{\psi _1 \psi _2} v^2$, $\mathcal{M}_{46}= \left(\lambda _{\eta \psi _1}+2 \lambda _{\eta  \psi _1 \psi _2}\right)u v $, $\mathcal{M}_{54}=\lambda _{\eta  \psi _1 \psi _2} u^2+\mu _{\psi _1 \psi _2}^2+ \lambda _{\psi _1 \psi _2} v^2$, $\mathcal{M}_{55}=-\lambda _{\eta  \psi _1 \psi _2} u^2-\mu _{\psi _1 \psi _2}^2+\frac{1}{12}  \lambda _{\psi _2} v^2$, $\mathcal{M}_{56}= \left(\lambda _{\eta \psi _2}+2 \lambda _{\eta  \psi _1 \psi _2}\right)  u v$.

{	
	The CP-odd particles ($\phi_h^0, \phi_{2h}^0$ and $ \chi^I$) mass matrix is given by
	\begin{eqnarray}
		\mathcal{M}^2_{CP-odd}&=	\left(
		\begin{array}{ccc}
			\frac{v_{2 h} \mu _{\phi _1 \phi _2}^2}{v_h} & -\mu _{\phi _1 \phi _2}^2 & 0 \\
			-\mu _{\phi _1 \phi _2}^2 & \frac{v_h \mu _{\phi _1 \phi _2}^2}{v_{2 h}} & 0 \\
			0 & 0 & 2 \mu _{\chi ^i}^2 \\
		\end{array}
		\right)
	\end{eqnarray}
	Using the same minimization we can also get the charged scalar ($\phi_h^\pm $ and $\phi_{2h}^\pm$) mass matrix as
	\begin{eqnarray}
		\mathcal{M}^2_{Charged}&=\frac{1}{4}\, \left(
		\begin{array}{cc}
			v_{2 h} (\lambda _{4 h} v_{2 h} + \frac{2 \mu _{\phi _1 \phi _2}^2}{v_h}) & -(2 \mu _{\phi _1 \phi _2}^2+  \lambda _{4 h} v_h v_{2 h}) \\
			-(2 \mu _{\phi _1 \phi _2}^2+  \lambda _{4 h}  v_h v_{2 h})& v_h ( \lambda _{4 h} v_h +\frac{2 \mu _{\phi _1 \phi _2}^2}{v_{2 h}} ) \\
		\end{array}
		\right)
	\end{eqnarray}
	One of the eigenvalues of these mass matrices is zero corresponds to the neutral and charged Goldstone Bosons.}                                                                                                                                                                                                                                                                                                                                                                                                                                                                                                                                                                                                                                                                                                                                                                                                                                                                                                                                                                                                                                                                                                                                                                                                                                                                                                                                                                                                                                                                                                                                                                                                                                                                                                                                                                                                                                                                                                     
We considered the VEV of $\phi_2$ is very small $v_{2h}\sim 0.1$ eV and the other singlet scalar VEVs are considered as $v\sim 10^{12}$ GeV, $u\sim10^{10}$ GeV and $v_{\chi}\sim 1000$ GeV. Hence we need a negative $\mu _{\phi _1 \phi _2}^2$ with $\lambda_{2h}=\mathcal{O}(0.1)$ to get positive $2 \lambda _{2 h} v_{2 h}^2-\frac{v_h \mu _{\phi _1 \phi _2}^2}{v_{2 h}}$. One can see that for the choice of these VEVs, the off-diagonal second-row and second column become too small as compared to the $\mathcal{M}^2_{CP-even} (2,2)$ component $2 \lambda _{2 h} v_{2 h}^2-\frac{v_h \mu _{\phi _1 \phi _2}^2}{v_{2 h}} >> 100 $ TeV. The mixing between $\phi_{2h}$ scalar field to the other sectors is too small. Hence, it remains decoupled from the other sectors. The CP-even scalar field of SM like doublet $\phi_h$ can have huge mixing with the other sector as  large VEVs ($v$ and $u$) are situated in the off-diagonal mass matrix. It is also true for the CP-even part of the complex scalar singlet $\chi$. These two sectors can be completely decoupled from the other scalar sectors. We also check that for $\lambda_{\zeta \phi_1}, \lambda_{\chi \zeta } = \mathcal{O}(0.1)$ the mixing remains too small $\sim \frac{v_h}{v} \lesssim 10^{-8}$.
Hence the above CP-even mass matrix in eqn.~\ref{eq:cpevenmass}, now can be decoupled in three sectors as
\begin{eqnarray}
	&&		\mathcal{M}^2 (\phi_{2h}) \approx 2 \lambda _{2 h} v_{2 h}^2 -\frac{v_h \mu _{\phi _1 \phi _2}^2}{v_{2 h}},\,\, \mathcal{M}^2 (\phi_{h},\chi ) = \left(
	\begin{array}{cc}
		2  \lambda _h v_h^2& \lambda _{\chi \phi _1} v_h v_{\chi } \\
		\lambda _{\chi \phi _1}v_h v_{\chi }  & \frac{1}{12} \lambda _{\chi }  v_{\chi }^2\\
	\end{array}
	\right)\\
	&&\nn	\mathcal{M}^2 (\eta,\psi_1,\psi_2) ={ \left(
		\begin{array}{ccc}
			-\lambda _{\eta  \psi _1 \psi _2} u^2-\mu _{\psi _1 \psi _2}^2+\frac{1}{12} v^2 \lambda _{\psi _1} & \lambda _{\eta  \psi _1 \psi _2} u^2+\mu _{\psi _1 \psi _2}^2+v^2 \lambda _{\psi _1 \psi _2} & u v \left(\lambda _{\eta \psi _1}+2 \lambda _{\eta  \psi _1 \psi _2}\right) \\
			\lambda _{\eta  \psi _1 \psi _2} u^2+\mu _{\psi _1 \psi _2}^2+v^2 \lambda _{\psi _1 \psi _2} & -\lambda _{\eta  \psi _1 \psi _2} u^2-\mu _{\psi _1 \psi _2}^2+\frac{1}{12} v^2 \lambda _{\psi _2} & u v \left(\lambda _{\eta \psi _2}+2 \lambda _{\eta  \psi _1 \psi _2}\right) \\
			u v \left(\lambda _{\eta \psi _1}+2 \lambda _{\eta  \psi _1 \psi _2}\right) & u v \left(\lambda _{\eta \psi _2}+2 \lambda _{\eta  \psi _1 \psi _2}\right) & \frac{u^2 \lambda _{\eta }}{12} \\
		\end{array}
		\right) }
\end{eqnarray}

The CP-odd and charged particles also remain decoupled using the above conditions. Hence the at electroweak scale the decoupled scalar potential can be written as 
\begin{eqnarray}
	V &= &-\mu_{\phi_1}^2 \phi_1^{\dagger}\phi_1 + \lambda_1 (\phi_1^{\dagger}\phi_1)^2 + \lambda_2 (\phi_1^{\dagger}\phi_1) (\chi^{\dagger} \chi ) \nn \\
	&&- \frac{1}{2} \mu_{\chi^R}^2 \chi^{\dagger} \chi  - \frac{1}{2}  \mu_{\chi^I}^2 (\chi^2 + h.c) + 
	\frac{\lambda_3}{4!} (\chi^{\dagger} \chi )^2.
\end{eqnarray}
For simplicity, we change the notation of quartic couplings as $\lambda_h\equiv \lambda_1$, $\lambda_{\chi \phi_1} \equiv \lambda_2$ and $\lambda_\chi \equiv \lambda_3$ respectively.

\bibliographystyle{utphys}
\bibliography{txbib}

\providecommand{\href}[2]{#2}\begingroup\raggedright\begin{thebibliography}{100}

\bibitem{Chatrchyan:2012xdj}
{\bfseries CMS} Collaboration, S.~Chatrchyan {\em et~al.}, ``{Observation of a
  new boson at a mass of 125 GeV with the CMS experiment at the LHC},''
  \href{http://dx.doi.org/10.1016/j.physletb.2012.08.021}{{\em Phys. Lett.}
  {\bfseries B716} (2012) 30--61},
\href{http://arxiv.org/abs/1207.7235}{{\ttfamily arXiv:1207.7235 [hep-ex]}}.

\bibitem{Aad:2012tfa}
{\bfseries ATLAS} Collaboration, G.~Aad {\em et~al.}, ``{Observation of a new
  particle in the search for the Standard Model Higgs boson with the ATLAS
  detector at the LHC},''
  \href{http://dx.doi.org/10.1016/j.physletb.2012.08.020}{{\em Phys. Lett.}
  {\bfseries B716} (2012) 1--29},
\href{http://arxiv.org/abs/1207.7214}{{\ttfamily arXiv:1207.7214 [hep-ex]}}.

\bibitem{An:2012eh}
{\bfseries Daya Bay} Collaboration, F.~P. An {\em et~al.}, ``{Observation of
  electron-antineutrino disappearance at Daya Bay},''
  \href{http://dx.doi.org/10.1103/PhysRevLett.108.171803}{{\em Phys. Rev.
  Lett.} {\bfseries 108} (2012) 171803},
\href{http://arxiv.org/abs/1203.1669}{{\ttfamily arXiv:1203.1669 [hep-ex]}}.

\bibitem{Abe:2011fz}
{\bfseries Double Chooz} Collaboration, Y.~Abe {\em et~al.}, ``{Indication of
  Reactor $\bar{\nu}_e$ Disappearance in the Double Chooz Experiment},''
  \href{http://dx.doi.org/10.1103/PhysRevLett.108.131801}{{\em Phys. Rev.
  Lett.} {\bfseries 108} (2012) 131801},
\href{http://arxiv.org/abs/1112.6353}{{\ttfamily arXiv:1112.6353 [hep-ex]}}.

\bibitem{Abe:2016nxk}
{\bfseries Super-Kamiokande} Collaboration, K.~Abe {\em et~al.}, ``{Solar
  Neutrino Measurements in Super-Kamiokande-IV},''
  \href{http://dx.doi.org/10.1103/PhysRevD.94.052010}{{\em Phys. Rev.}
  {\bfseries D94} no.~5, (2016) 052010},
\href{http://arxiv.org/abs/1606.07538}{{\ttfamily arXiv:1606.07538 [hep-ex]}}.

\bibitem{Jungman:1995df}
G.~Jungman, M.~Kamionkowski, and K.~Griest, ``{Supersymmetric dark matter},''
  \href{http://dx.doi.org/10.1016/0370-1573(95)00058-5}{{\em Phys. Rept.}
  {\bfseries 267} (1996) 195--373},
  \href{http://arxiv.org/abs/hep-ph/9506380}{{\ttfamily arXiv:hep-ph/9506380}}.

\bibitem{Bertone:2004pz}
G.~Bertone, D.~Hooper, and J.~Silk, ``{Particle dark matter: Evidence,
  candidates and constraints},''
  \href{http://dx.doi.org/10.1016/j.physrep.2004.08.031}{{\em Phys. Rept.}
  {\bfseries 405} (2005) 279--390},
\href{http://arxiv.org/abs/hep-ph/0404175}{{\ttfamily arXiv:hep-ph/0404175
  [hep-ph]}}.

\bibitem{Athanassopoulos:1996ds}
{\bfseries LSND} Collaboration, C.~Athanassopoulos {\em et~al.}, ``{The Liquid
  scintillator neutrino detector and LAMPF neutrino source},''
  \href{http://dx.doi.org/10.1016/S0168-9002(96)01155-2}{{\em Nucl. Instrum.
  Meth.} {\bfseries A388} (1997) 149--172},
\href{http://arxiv.org/abs/nucl-ex/9605002}{{\ttfamily arXiv:nucl-ex/9605002
  [nucl-ex]}}.

\bibitem{Aguilar:2001ty}
{\bfseries LSND} Collaboration, A.~Aguilar-Arevalo {\em et~al.}, ``{Evidence
  for neutrino oscillations from the observation of $\bar{\nu}_e$ appearance in
  a $\bar{\nu}_\mu$ beam},''
  \href{http://dx.doi.org/10.1103/PhysRevD.64.112007}{{\em Phys. Rev. D}
  {\bfseries 64} (2001) 112007},
  \href{http://arxiv.org/abs/hep-ex/0104049}{{\ttfamily arXiv:hep-ex/0104049}}.

\bibitem{Aguilar-Arevalo:2018gpe}
{\bfseries MiniBooNE} Collaboration, A.~A. Aguilar-Arevalo {\em et~al.},
  ``{Significant Excess of ElectronLike Events in the MiniBooNE Short-Baseline
  Neutrino Experiment},''
  \href{http://dx.doi.org/10.1103/PhysRevLett.121.221801}{{\em Phys. Rev.
  Lett.} {\bfseries 121} no.~22, (2018) 221801},
\href{http://arxiv.org/abs/1805.12028}{{\ttfamily arXiv:1805.12028 [hep-ex]}}.

\bibitem{Hampel:1998xg}
{\bfseries GALLEX} Collaboration, W.~Hampel {\em et~al.}, ``{GALLEX solar
  neutrino observations: Results for GALLEX IV},''
  \href{http://dx.doi.org/10.1016/S0370-2693(98)01579-2}{{\em Phys. Lett. B}
  {\bfseries 447} (1999) 127--133}.

\bibitem{Abdurashitov:1999zd}
{\bfseries SAGE} Collaboration, J.~Abdurashitov {\em et~al.}, ``{Measurement of
  the solar neutrino capture rate with gallium metal},''
  \href{http://dx.doi.org/10.1103/PhysRevC.60.055801}{{\em Phys. Rev. C}
  {\bfseries 60} (1999) 055801},
  \href{http://arxiv.org/abs/astro-ph/9907113}{{\ttfamily
  arXiv:astro-ph/9907113}}.

\bibitem{Collin:2016rao}
G.~Collin, C.~Argüelles, J.~Conrad, and M.~Shaevitz, ``{Sterile Neutrino Fits
  to Short Baseline Data},''
  \href{http://dx.doi.org/10.1016/j.nuclphysb.2016.02.024}{{\em Nucl. Phys. B}
  {\bfseries 908} (2016) 354--365},
  \href{http://arxiv.org/abs/1602.00671}{{\ttfamily arXiv:1602.00671
  [hep-ph]}}.

\bibitem{Mention:2011rk}
G.~Mention, M.~Fechner, T.~Lasserre, T.~Mueller, D.~Lhuillier, M.~Cribier, and
  A.~Letourneau, ``{The Reactor Antineutrino Anomaly},''
  \href{http://dx.doi.org/10.1103/PhysRevD.83.073006}{{\em Phys. Rev. D}
  {\bfseries 83} (2011) 073006},
  \href{http://arxiv.org/abs/1101.2755}{{\ttfamily arXiv:1101.2755 [hep-ex]}}.

\bibitem{Mueller:2011nm}
T.~Mueller {\em et~al.}, ``{Improved Predictions of Reactor Antineutrino
  Spectra},'' \href{http://dx.doi.org/10.1103/PhysRevC.83.054615}{{\em Phys.
  Rev. C} {\bfseries 83} (2011) 054615},
  \href{http://arxiv.org/abs/1101.2663}{{\ttfamily arXiv:1101.2663 [hep-ex]}}.

\bibitem{Dodelson:1993je}
S.~Dodelson and L.~M. Widrow, ``{Sterile-neutrinos as dark matter},''
  \href{http://dx.doi.org/10.1103/PhysRevLett.72.17}{{\em Phys. Rev. Lett.}
  {\bfseries 72} (1994) 17--20},
\href{http://arxiv.org/abs/hep-ph/9303287}{{\ttfamily arXiv:hep-ph/9303287
  [hep-ph]}}.

\bibitem{Ruchayskiy:2012si}
O.~Ruchayskiy and A.~Ivashko, ``{Restrictions on the lifetime of sterile
  neutrinos from primordial nucleosynthesis},''
  \href{http://dx.doi.org/10.1088/1475-7516/2012/10/014}{{\em JCAP} {\bfseries
  1210} (2012) 014},
\href{http://arxiv.org/abs/1202.2841}{{\ttfamily arXiv:1202.2841 [hep-ph]}}.

\bibitem{Abazajian:2012ys}
K.~N. Abazajian {\em et~al.}, ``{Light Sterile Neutrinos: A White Paper},''
\href{http://arxiv.org/abs/1204.5379}{{\ttfamily arXiv:1204.5379 [hep-ph]}}.

\bibitem{Abazajian:2017tcc}
K.~N. Abazajian, ``{Sterile neutrinos in cosmology},''
  \href{http://dx.doi.org/10.1016/j.physrep.2017.10.003}{{\em Phys. Rept.}
  {\bfseries 711-712} (2017) 1--28},
\href{http://arxiv.org/abs/1705.01837}{{\ttfamily arXiv:1705.01837 [hep-ph]}}.

\bibitem{Adhikari:2016bei}
M.~Drewes {\em et~al.}, ``{A White Paper on keV Sterile Neutrino Dark
  Matter},'' \href{http://dx.doi.org/10.1088/1475-7516/2017/01/025}{{\em JCAP}
  {\bfseries 1701} no.~01, (2017) 025},
\href{http://arxiv.org/abs/1602.04816}{{\ttfamily arXiv:1602.04816 [hep-ph]}}.

\bibitem{Atre:2009rg}
A.~Atre, T.~Han, S.~Pascoli, and B.~Zhang, ``{The Search for Heavy Majorana
  Neutrinos},'' \href{http://dx.doi.org/10.1088/1126-6708/2009/05/030}{{\em
  JHEP} {\bfseries 05} (2009) 030},
  \href{http://arxiv.org/abs/0901.3589}{{\ttfamily arXiv:0901.3589 [hep-ph]}}.

\bibitem{Deppisch:2015qwa}
F.~F. Deppisch, P.~S. Bhupal~Dev, and A.~Pilaftsis, ``{Neutrinos and Collider
  Physics},'' \href{http://dx.doi.org/10.1088/1367-2630/17/7/075019}{{\em New
  J. Phys.} {\bfseries 17} no.~7, (2015) 075019},
  \href{http://arxiv.org/abs/1502.06541}{{\ttfamily arXiv:1502.06541
  [hep-ph]}}.

\bibitem{Aker:2019uuj}
{\bfseries KATRIN} Collaboration, M.~Aker {\em et~al.}, ``{Improved Upper Limit
  on the Neutrino Mass from a Direct Kinematic Method by KATRIN},''
  \href{http://dx.doi.org/10.1103/PhysRevLett.123.221802}{{\em Phys. Rev.
  Lett.} {\bfseries 123} no.~22, (2019) 221802},
  \href{http://arxiv.org/abs/1909.06048}{{\ttfamily arXiv:1909.06048
  [hep-ex]}}.

\bibitem{Hinshaw:2012aka}
{\bfseries WMAP} Collaboration, G.~Hinshaw {\em et~al.}, ``{Nine-Year Wilkinson
  Microwave Anisotropy Probe (WMAP) Observations: Cosmological Parameter
  Results},'' \href{http://dx.doi.org/10.1088/0067-0049/208/2/19}{{\em
  Astrophys. J. Suppl.} {\bfseries 208} (2013) 19},
  \href{http://arxiv.org/abs/1212.5226}{{\ttfamily arXiv:1212.5226
  [astro-ph.CO]}}.

\bibitem{Ade:2015xua}
{\bfseries Planck} Collaboration, P.~Ade {\em et~al.}, ``{Planck 2015 results.
  XIII. Cosmological parameters},''
  \href{http://dx.doi.org/10.1051/0004-6361/201525830}{{\em Astron. Astrophys.}
  {\bfseries 594} (2016) A13},
  \href{http://arxiv.org/abs/1502.01589}{{\ttfamily arXiv:1502.01589
  [astro-ph.CO]}}.

\bibitem{Giunti:2019aiy}
C.~Giunti and T.~Lasserre, ``{eV-scale Sterile Neutrinos},''
  \href{http://dx.doi.org/10.1146/annurev-nucl-101918-023755}{{\em Ann. Rev.
  Nucl. Part. Sci.} {\bfseries 69} (2019) 163--190},
  \href{http://arxiv.org/abs/1901.08330}{{\ttfamily arXiv:1901.08330
  [hep-ph]}}.

\bibitem{Boser:2019rta}
S.~B\"oser, C.~Buck, C.~Giunti, J.~Lesgourgues, L.~Ludhova, S.~Mertens,
  A.~Schukraft, and M.~Wurm, ``{Status of Light Sterile Neutrino Searches},''
  \href{http://dx.doi.org/10.1016/j.ppnp.2019.103736}{{\em Prog. Part. Nucl.
  Phys.} {\bfseries 111} (2020) 103736},
  \href{http://arxiv.org/abs/1906.01739}{{\ttfamily arXiv:1906.01739
  [hep-ex]}}.

\bibitem{Laine:2008pg}
M.~Laine and M.~Shaposhnikov, ``{Sterile neutrino dark matter as a consequence
  of nuMSM-induced lepton asymmetry},''
  \href{http://dx.doi.org/10.1088/1475-7516/2008/06/031}{{\em JCAP} {\bfseries
  0806} (2008) 031},
\href{http://arxiv.org/abs/0804.4543}{{\ttfamily arXiv:0804.4543 [hep-ph]}}.

\bibitem{Abada:2014zra}
A.~Abada, G.~Arcadi, and M.~Lucente, ``{Dark Matter in the minimal Inverse
  Seesaw mechanism},''
  \href{http://dx.doi.org/10.1088/1475-7516/2014/10/001}{{\em JCAP} {\bfseries
  1410} (2014) 001},
\href{http://arxiv.org/abs/1406.6556}{{\ttfamily arXiv:1406.6556 [hep-ph]}}.

\bibitem{Barry:2011wb}
J.~Barry, W.~Rodejohann, and H.~Zhang, ``{Light Sterile Neutrinos: Models and
  Phenomenology},'' \href{http://dx.doi.org/10.1007/JHEP07(2011)091}{{\em JHEP}
  {\bfseries 07} (2011) 091},
\href{http://arxiv.org/abs/1105.3911}{{\ttfamily arXiv:1105.3911 [hep-ph]}}.

\bibitem{Zhang:2011vh}
H.~Zhang, ``{Light Sterile Neutrino in the Minimal Extended Seesaw},''
  \href{http://dx.doi.org/10.1016/j.physletb.2012.06.074}{{\em Phys. Lett.}
  {\bfseries B714} (2012) 262--266},
\href{http://arxiv.org/abs/1110.6838}{{\ttfamily arXiv:1110.6838 [hep-ph]}}.

\bibitem{Nath:2016mts}
N.~Nath, M.~Ghosh, S.~Goswami, and S.~Gupta, ``{Phenomenological study of
  extended seesaw model for light sterile neutrino},''
  \href{http://dx.doi.org/10.1007/JHEP03(2017)075}{{\em JHEP} {\bfseries 03}
  (2017) 075},
\href{http://arxiv.org/abs/1610.09090}{{\ttfamily arXiv:1610.09090 [hep-ph]}}.

\bibitem{Das:2018qyt}
P.~Das, A.~Mukherjee, and M.~K. Das, ``{Active and sterile neutrino
  phenomenology with $A_4$ based minimal extended seesaw},''
  \href{http://dx.doi.org/10.1016/j.nuclphysb.2019.02.024}{{\em Nucl. Phys.}
  {\bfseries B941} (2019) 755--779},
\href{http://arxiv.org/abs/1805.09231}{{\ttfamily arXiv:1805.09231 [hep-ph]}}.

\bibitem{BhupalDev:2012jvh}
P.~S. Bhupal~Dev and A.~Pilaftsis, ``{Light and Superlight Sterile Neutrinos in
  the Minimal Radiative Inverse Seesaw Model},''
  \href{http://dx.doi.org/10.1103/PhysRevD.87.053007}{{\em Phys. Rev. D}
  {\bfseries 87} no.~5, (2013) 053007},
  \href{http://arxiv.org/abs/1212.3808}{{\ttfamily arXiv:1212.3808 [hep-ph]}}.

\bibitem{Das:2019kmn}
P.~Das and M.~K. Das, ``{Phenomenology of $keV$ sterile neutrino in minimal
  extended seesaw},'' \href{http://dx.doi.org/10.1142/S0217751X20501250}{{\em
  Int. J. Mod. Phys. A} {\bfseries 35} no.~22, (2020) 2050125},
  \href{http://arxiv.org/abs/1908.08417}{{\ttfamily arXiv:1908.08417
  [hep-ph]}}.

\bibitem{Benso:2019jog}
C.~Benso, V.~Brdar, M.~Lindner, and W.~Rodejohann, ``{Prospects for Finding
  Sterile Neutrino Dark Matter at KATRIN},''
  \href{http://dx.doi.org/10.1103/PhysRevD.100.115035}{{\em Phys. Rev.}
  {\bfseries D100} no.~11, (2019) 115035},
\href{http://arxiv.org/abs/1911.00328}{{\ttfamily arXiv:1911.00328 [hep-ph]}}.

\bibitem{Kageyama:2002zw}
A.~Kageyama, S.~Kaneko, N.~Shimoyama, and M.~Tanimoto, ``{Seesaw realization of
  the texture zeros in the neutrino mass matrix},''
  \href{http://dx.doi.org/10.1016/S0370-2693(02)01964-0}{{\em Phys. Lett. B}
  {\bfseries 538} (2002) 96--106},
  \href{http://arxiv.org/abs/hep-ph/0204291}{{\ttfamily arXiv:hep-ph/0204291}}.

\bibitem{Zhang:2013mb}
Y.~Zhang, ``{Majorana neutrino mass matrices with three texture zeros and the
  sterile neutrino},'' \href{http://dx.doi.org/10.1103/PhysRevD.87.053020}{{\em
  Phys. Rev. D} {\bfseries 87} no.~5, (2013) 053020},
  \href{http://arxiv.org/abs/1301.7302}{{\ttfamily arXiv:1301.7302 [hep-ph]}}.

\bibitem{Borah:2016xkc}
D.~Borah, M.~Ghosh, S.~Gupta, S.~Prakash, and S.~K. Raut, ``{Analysis of
  four-zero textures in the $3+1$ neutrino framework},''
  \href{http://dx.doi.org/10.1103/PhysRevD.94.113001}{{\em Phys. Rev. D}
  {\bfseries 94} no.~11, (2016) 113001},
  \href{http://arxiv.org/abs/1606.02076}{{\ttfamily arXiv:1606.02076
  [hep-ph]}}.

\bibitem{Borah:2015vra}
M.~Borah, D.~Borah, and M.~K. Das, ``{Discriminating Majorana neutrino textures
  in light of the baryon asymmetry},''
  \href{http://dx.doi.org/10.1103/PhysRevD.91.113008}{{\em Phys. Rev. D}
  {\bfseries 91} (2015) 113008},
  \href{http://arxiv.org/abs/1503.03431}{{\ttfamily arXiv:1503.03431
  [hep-ph]}}.

\bibitem{Babu:2009fd}
K.~S. Babu, \href{http://dx.doi.org/10.1142/9789812838360_0002}{``{TASI
  Lectures on Flavor Physics},''} in {\em {Proceedings of Theoretical Advanced
  Study Institute in Elementary Particle Physics on The dawn of the LHC era
  (TASI 2008): Boulder, USA, June 2-27, 2008}}, pp.~49--123.
\newblock 2010.
\newblock
\href{http://arxiv.org/abs/0910.2948}{{\ttfamily arXiv:0910.2948 [hep-ph]}}.
\newblock

\bibitem{Dziewit:2016qri}
B.~Dziewit, J.~Holeczek, M.~Richter, S.~Zajk~ac, and M.~Zralek, ``{Texture
  zeros in neutrino mass matrix},''
  \href{http://dx.doi.org/10.1134/S1063778817020144}{{\em Phys. Atom. Nucl.}
  {\bfseries 80} no.~2, (2017) 353--357},
  \href{http://arxiv.org/abs/1605.06547}{{\ttfamily arXiv:1605.06547
  [hep-ph]}}.

\bibitem{Borah:2017azf}
D.~Borah, M.~Ghosh, S.~Gupta, and S.~K. Raut, ``{Texture zeros of low-energy
  Majorana neutrino mass matrix in 3+1 scheme},''
  \href{http://dx.doi.org/10.1103/PhysRevD.96.055017}{{\em Phys. Rev. D}
  {\bfseries 96} no.~5, (2017) 055017},
  \href{http://arxiv.org/abs/1706.02017}{{\ttfamily arXiv:1706.02017
  [hep-ph]}}.

\bibitem{Sarma:2018bgf}
N.~Sarma, K.~Bora, and D.~Borah, ``{Compatibility of $A_{4}$ Flavour Symmetric
  Minimal Extended Seesaw with $(3+1)$ Neutrino Data},''
  \href{http://dx.doi.org/10.1140/epjc/s10052-019-6584-z}{{\em Eur. Phys. J. C}
  {\bfseries 79} no.~2, (2019) 129},
  \href{http://arxiv.org/abs/1810.05826}{{\ttfamily arXiv:1810.05826
  [hep-ph]}}.

\bibitem{Davidson:2008bu}
S.~Davidson, E.~Nardi, and Y.~Nir, ``{Leptogenesis},''
  \href{http://dx.doi.org/10.1016/j.physrep.2008.06.002}{{\em Phys. Rept.}
  {\bfseries 466} (2008) 105--177},
\href{http://arxiv.org/abs/0802.2962}{{\ttfamily arXiv:0802.2962 [hep-ph]}}.

\bibitem{Buchmuller:2004tu}
W.~Buchmuller, P.~Di~Bari, and M.~Plumacher, ``{Some aspects of thermal
  leptogenesis},'' \href{http://dx.doi.org/10.1088/1367-2630/6/1/105}{{\em New
  J. Phys.} {\bfseries 6} (2004) 105},
\href{http://arxiv.org/abs/hep-ph/0406014}{{\ttfamily arXiv:hep-ph/0406014
  [hep-ph]}}.

\bibitem{Frossard:2012pc}
T.~Frossard, M.~Garny, A.~Hohenegger, A.~Kartavtsev, and D.~Mitrouskas,
  ``{Systematic approach to thermal leptogenesis},''
  \href{http://dx.doi.org/10.1103/PhysRevD.87.085009}{{\em Phys. Rev.}
  {\bfseries D87} no.~8, (2013) 085009},
\href{http://arxiv.org/abs/1211.2140}{{\ttfamily arXiv:1211.2140 [hep-ph]}}.

\bibitem{Joshipura:2001ya}
A.~S. Joshipura, E.~A. Paschos, and W.~Rodejohann, ``{Leptogenesis in
  left-right symmetric theories},''
  \href{http://dx.doi.org/10.1016/S0550-3213(01)00346-7}{{\em Nucl. Phys. B}
  {\bfseries 611} (2001) 227--238},
  \href{http://arxiv.org/abs/hep-ph/0104228}{{\ttfamily arXiv:hep-ph/0104228}}.

\bibitem{Sakharov:1967dj}
A.~D. Sakharov, ``{Violation of CP Invariance, C asymmetry, and baryon
  asymmetry of the universe},''
  \href{http://dx.doi.org/10.1070/PU1991v034n05ABEH002497}{{\em Pisma Zh. Eksp.
  Teor. Fiz.} {\bfseries 5} (1967) 32--35}.
[Usp. Fiz. Nauk161,no.5,61(1991)].

\bibitem{Pilaftsis:2003gt}
A.~Pilaftsis and T.~E. Underwood, ``{Resonant leptogenesis},''
  \href{http://dx.doi.org/10.1016/j.nuclphysb.2004.05.029}{{\em Nucl. Phys. B}
  {\bfseries 692} (2004) 303--345},
  \href{http://arxiv.org/abs/hep-ph/0309342}{{\ttfamily arXiv:hep-ph/0309342}}.

\bibitem{Hambye:2001eu}
T.~Hambye, ``{Leptogenesis at the TeV scale},''
  \href{http://dx.doi.org/10.1016/S0550-3213(02)00293-6}{{\em Nucl. Phys. B}
  {\bfseries 633} (2002) 171--192},
  \href{http://arxiv.org/abs/hep-ph/0111089}{{\ttfamily arXiv:hep-ph/0111089}}.

\bibitem{Cirigliano:2006nu}
V.~Cirigliano, G.~Isidori, and V.~Porretti, ``{CP violation and Leptogenesis in
  models with Minimal Lepton Flavour Violation},''
  \href{http://dx.doi.org/10.1016/j.nuclphysb.2006.11.015}{{\em Nucl. Phys. B}
  {\bfseries 763} (2007) 228--246},
  \href{http://arxiv.org/abs/hep-ph/0607068}{{\ttfamily arXiv:hep-ph/0607068}}.

\bibitem{Chun:2007vh}
E.~Chun and K.~Turzynski, ``{Quasi-degenerate neutrinos and leptogenesis from
  L(mu) - L(tau)},'' \href{http://dx.doi.org/10.1103/PhysRevD.76.053008}{{\em
  Phys. Rev. D} {\bfseries 76} (2007) 053008},
  \href{http://arxiv.org/abs/hep-ph/0703070}{{\ttfamily arXiv:hep-ph/0703070}}.

\bibitem{Kitabayashi:2007bs}
T.~Kitabayashi, ``{Remark on the minimal seesaw model and leptogenesis with
  tri/bi-maximal mixing},''
  \href{http://dx.doi.org/10.1103/PhysRevD.76.033002}{{\em Phys. Rev. D}
  {\bfseries 76} (2007) 033002},
  \href{http://arxiv.org/abs/hep-ph/0703303}{{\ttfamily arXiv:hep-ph/0703303}}.

\bibitem{Bambhaniya:2016rbb}
G.~Bambhaniya, P.~Bhupal~Dev, S.~Goswami, S.~Khan, and W.~Rodejohann,
  ``{Naturalness, Vacuum Stability and Leptogenesis in the Minimal Seesaw
  Model},'' \href{http://dx.doi.org/10.1103/PhysRevD.95.095016}{{\em Phys. Rev.
  D} {\bfseries 95} no.~9, (2017) 095016},
  \href{http://arxiv.org/abs/1611.03827}{{\ttfamily arXiv:1611.03827
  [hep-ph]}}.

\bibitem{Liu:1993tg}
J.~Liu and G.~Segre, ``{Reexamination of generation of baryon and lepton number
  asymmetries by heavy particle decay},''
  \href{http://dx.doi.org/10.1103/PhysRevD.48.4609}{{\em Phys. Rev. D}
  {\bfseries 48} (1993) 4609--4612},
  \href{http://arxiv.org/abs/hep-ph/9304241}{{\ttfamily arXiv:hep-ph/9304241}}.

\bibitem{Iso:2013lba}
S.~Iso, K.~Shimada, and M.~Yamanaka, ``{Kadanoff-Baym approach to the thermal
  resonant leptogenesis},''
  \href{http://dx.doi.org/10.1007/JHEP04(2014)062}{{\em JHEP} {\bfseries 04}
  (2014) 062}, \href{http://arxiv.org/abs/1312.7680}{{\ttfamily arXiv:1312.7680
  [hep-ph]}}.

\bibitem{kadanoff2018quantum}
L.~P. Kadanoff, {\em Quantum statistical mechanics}.
\newblock CRC Press, 2018.

\bibitem{Garny:2011hg}
M.~Garny, A.~Kartavtsev, and A.~Hohenegger, ``{Leptogenesis from first
  principles in the resonant regime},''
  \href{http://dx.doi.org/10.1016/j.aop.2012.10.007}{{\em Annals Phys.}
  {\bfseries 328} (2013) 26--63},
  \href{http://arxiv.org/abs/1112.6428}{{\ttfamily arXiv:1112.6428 [hep-ph]}}.

\bibitem{BhupalDev:2014oar}
P.~S. Bhupal~Dev, P.~Millington, A.~Pilaftsis, and D.~Teresi,
  ``{Kadanoff\textendash{}Baym approach to flavour mixing and oscillations in
  resonant leptogenesis},''
  \href{http://dx.doi.org/10.1016/j.nuclphysb.2014.12.003}{{\em Nucl. Phys. B}
  {\bfseries 891} (2015) 128--158},
  \href{http://arxiv.org/abs/1410.6434}{{\ttfamily arXiv:1410.6434 [hep-ph]}}.

\bibitem{Kartavtsev:2015vto}
A.~Kartavtsev, P.~Millington, and H.~Vogel, ``{Lepton asymmetry from mixing and
  oscillations},'' \href{http://dx.doi.org/10.1007/JHEP06(2016)066}{{\em JHEP}
  {\bfseries 06} (2016) 066}, \href{http://arxiv.org/abs/1601.03086}{{\ttfamily
  arXiv:1601.03086 [hep-ph]}}.

\bibitem{Millington:2012pf}
P.~Millington and A.~Pilaftsis, ``{Perturbative nonequilibrium thermal field
  theory},'' \href{http://dx.doi.org/10.1103/PhysRevD.88.085009}{{\em Phys.
  Rev. D} {\bfseries 88} no.~8, (2013) 085009},
  \href{http://arxiv.org/abs/1211.3152}{{\ttfamily arXiv:1211.3152 [hep-ph]}}.

\bibitem{Dev:2014laa}
P.~Bhupal~Dev, P.~Millington, A.~Pilaftsis, and D.~Teresi, ``{Flavour Covariant
  Transport Equations: an Application to Resonant Leptogenesis},''
  \href{http://dx.doi.org/10.1016/j.nuclphysb.2014.06.020}{{\em Nucl. Phys. B}
  {\bfseries 886} (2014) 569--664},
  \href{http://arxiv.org/abs/1404.1003}{{\ttfamily arXiv:1404.1003 [hep-ph]}}.

\bibitem{DeSimone:2007edo}
A.~De~Simone and A.~Riotto, ``{On Resonant Leptogenesis},''
  \href{http://dx.doi.org/10.1088/1475-7516/2007/08/013}{{\em JCAP} {\bfseries
  08} (2007) 013}, \href{http://arxiv.org/abs/0705.2183}{{\ttfamily
  arXiv:0705.2183 [hep-ph]}}.

\bibitem{Deppisch:2010fr}
F.~F. Deppisch and A.~Pilaftsis, ``{Lepton Flavour Violation and theta(13) in
  Minimal Resonant Leptogenesis},''
  \href{http://dx.doi.org/10.1103/PhysRevD.83.076007}{{\em Phys. Rev. D}
  {\bfseries 83} (2011) 076007},
  \href{http://arxiv.org/abs/1012.1834}{{\ttfamily arXiv:1012.1834 [hep-ph]}}.

\bibitem{Cline:2006ts}
J.~M. Cline, ``{Baryogenesis},'' in {\em {Les Houches Summer School - Session
  86: Particle Physics and Cosmology: The Fabric of Spacetime Les Houches,
  France, July 31-August 25, 2006}}.
\newblock 2006.
\newblock
\href{http://arxiv.org/abs/hep-ph/0609145}{{\ttfamily arXiv:hep-ph/0609145
  [hep-ph]}}.
\newblock

\bibitem{Bilenky:2004wn}
S.~M. Bilenky, A.~Faessler, and F.~Simkovic, ``{The Majorana neutrino masses,
  neutrinoless double beta decay and nuclear matrix elements},''
  \href{http://dx.doi.org/10.1103/PhysRevD.70.033003}{{\em Phys. Rev.}
  {\bfseries D70} (2004) 033003},
\href{http://arxiv.org/abs/hep-ph/0402250}{{\ttfamily arXiv:hep-ph/0402250
  [hep-ph]}}.

\bibitem{Bilenky:2012qi}
S.~M. Bilenky and C.~Giunti, ``{Neutrinoless double-beta decay: A brief
  review},'' \href{http://dx.doi.org/10.1142/S0217732312300157}{{\em Mod. Phys.
  Lett.} {\bfseries A27} (2012) 1230015},
\href{http://arxiv.org/abs/1203.5250}{{\ttfamily arXiv:1203.5250 [hep-ph]}}.

\bibitem{Agostini:2018tnm}
{\bfseries GERDA} Collaboration, M.~Agostini {\em et~al.}, ``{Improved Limit on
  Neutrinoless Double-$\beta$ Decay of $^{76}$Ge from GERDA Phase II},''
  \href{http://dx.doi.org/10.1103/PhysRevLett.120.132503}{{\em Phys. Rev.
  Lett.} {\bfseries 120} no.~13, (2018) 132503},
\href{http://arxiv.org/abs/1803.11100}{{\ttfamily arXiv:1803.11100 [nucl-ex]}}.

\bibitem{Borgohain:2017akh}
H.~Borgohain and M.~K. Das, ``{Lepton number violation, lepton flavor
  violation, and baryogenesis in left-right symmetric model},''
  \href{http://dx.doi.org/10.1103/PhysRevD.96.075021}{{\em Phys. Rev.}
  {\bfseries D96} no.~7, (2017) 075021},
\href{http://arxiv.org/abs/1709.09542}{{\ttfamily arXiv:1709.09542 [hep-ph]}}.

\bibitem{Awasthi:2013ff}
R.~L. Awasthi, M.~K. Parida, and S.~Patra, ``{Neutrino masses, dominant
  neutrinoless double beta decay, and observable lepton flavor violation in
  left-right models and SO(10) grand unification with low mass $ W_R, Z_R$
  bosons},'' \href{http://dx.doi.org/10.1007/JHEP08(2013)122}{{\em JHEP}
  {\bfseries 08} (2013) 122}, \href{http://arxiv.org/abs/1302.0672}{{\ttfamily
  arXiv:1302.0672 [hep-ph]}}.

\bibitem{Abada:2018qok}
A.~Abada, A.~Hern\'andez-Cabezudo, and X.~Marcano, ``{Beta and Neutrinoless
  Double Beta Decays with KeV Sterile Fermions},''
  \href{http://dx.doi.org/10.1007/JHEP01(2019)041}{{\em JHEP} {\bfseries 01}
  (2019) 041}, \href{http://arxiv.org/abs/1807.01331}{{\ttfamily
  arXiv:1807.01331 [hep-ph]}}.

\bibitem{Hambye:2009pw}
T.~Hambye, F.~S. Ling, L.~Lopez~Honorez, and J.~Rocher, ``{Scalar Multiplet
  Dark Matter},'' \href{http://dx.doi.org/10.1007/JHEP05(2010)066,
  10.1088/1126-6708/2009/07/090}{{\em JHEP} {\bfseries 07} (2009) 090},
  \href{http://arxiv.org/abs/0903.4010}{{\ttfamily arXiv:0903.4010 [hep-ph]}}.
[Erratum: JHEP05,066(2010)].

\bibitem{Magana:2012ph}
J.~Magana and T.~Matos, ``{A brief Review of the Scalar Field Dark Matter
  model},'' \href{http://dx.doi.org/10.1088/1742-6596/378/1/012012}{{\em J.
  Phys. Conf. Ser.} {\bfseries 378} (2012) 012012},
\href{http://arxiv.org/abs/1201.6107}{{\ttfamily arXiv:1201.6107
  [astro-ph.CO]}}.

\bibitem{Khan:2017xyh}
N.~Khan, {\em {Exploring Extensions of the Scalar Sector of the Standard
  Model}}.
\newblock PhD thesis, Indian Inst. Tech., Indore, 2017.
\newblock
\href{http://arxiv.org/abs/1701.02205}{{\ttfamily arXiv:1701.02205 [hep-ph]}}.
\newblock

\bibitem{Das:2014fea}
D.~Das and U.~K. Dey, ``{Analysis of an extended scalar sector with $S_3$
  symmetry},'' \href{http://dx.doi.org/10.1103/PhysRevD.91.039905,
  10.1103/PhysRevD.89.095025}{{\em Phys. Rev.} {\bfseries D89} no.~9, (2014)
  095025}, \href{http://arxiv.org/abs/1404.2491}{{\ttfamily arXiv:1404.2491
  [hep-ph]}}.
[Erratum: Phys. Rev.D91,no.3,039905(2015)].

\bibitem{Das:2019ntw}
P.~Das, M.~K. Das, and N.~Khan, ``{Phenomenological study of neutrino mass,
  dark matter and baryogenesis within the framework of minimal extended
  seesaw},'' \href{http://dx.doi.org/10.1007/JHEP03(2020)018}{{\em JHEP}
  {\bfseries 03} (2020) 018}, \href{http://arxiv.org/abs/1911.07243}{{\ttfamily
  arXiv:1911.07243 [hep-ph]}}.

\bibitem{Esteban:2020cvm}
I.~Esteban, M.~Gonzalez-Garcia, M.~Maltoni, T.~Schwetz, and A.~Zhou, ``{The
  fate of hints: updated global analysis of three-flavor neutrino
  oscillations},'' \href{http://dx.doi.org/10.1007/JHEP09(2020)178}{{\em JHEP}
  {\bfseries 09} (2020) 178}, \href{http://arxiv.org/abs/2007.14792}{{\ttfamily
  arXiv:2007.14792 [hep-ph]}}.

\bibitem{Ghosh:2013nya}
M.~Ghosh, S.~Goswami, S.~Gupta, and C.~S. Kim, ``{Implication of a vanishing
  element in the 3+1 scenario},''
  \href{http://dx.doi.org/10.1103/PhysRevD.88.033009}{{\em Phys. Rev. D}
  {\bfseries 88} no.~3, (2013) 033009},
  \href{http://arxiv.org/abs/1305.0180}{{\ttfamily arXiv:1305.0180 [hep-ph]}}.

\bibitem{Ma:2009wi}
E.~Ma, ``{Neutrino Tribimaximal Mixing from A(4) Alone},''
  \href{http://dx.doi.org/10.1142/S021773231003361X}{{\em Mod. Phys. Lett.}
  {\bfseries A25} (2010) 2215--2221},
\href{http://arxiv.org/abs/0908.3165}{{\ttfamily arXiv:0908.3165 [hep-ph]}}.

\bibitem{Altarelli:2005yp}
G.~Altarelli and F.~Feruglio, ``{Tri-bimaximal neutrino mixing from discrete
  symmetry in extra dimensions},''
  \href{http://dx.doi.org/10.1016/j.nuclphysb.2005.05.005}{{\em Nucl. Phys.}
  {\bfseries B720} (2005) 64--88},
\href{http://arxiv.org/abs/hep-ph/0504165}{{\ttfamily arXiv:hep-ph/0504165
  [hep-ph]}}.

\bibitem{Mukherjee:2017pzq}
A.~Mukherjee, D.~Borah, and M.~K. Das, ``{Common Origin of Non-zero
  $\theta_{13}$ and Dark Matter in an $S_4$ Flavour Symmetric Model with
  Inverse Seesaw},'' \href{http://dx.doi.org/10.1103/PhysRevD.96.015014}{{\em
  Phys. Rev.} {\bfseries D96} no.~1, (2017) 015014},
\href{http://arxiv.org/abs/1703.06750}{{\ttfamily arXiv:1703.06750 [hep-ph]}}.

\bibitem{Davidson:2009ha}
S.~M. Davidson and H.~E. Logan, ``{Dirac neutrinos from a second Higgs
  doublet},'' \href{http://dx.doi.org/10.1103/PhysRevD.80.095008}{{\em Phys.
  Rev. D} {\bfseries 80} (2009) 095008},
  \href{http://arxiv.org/abs/0906.3335}{{\ttfamily arXiv:0906.3335 [hep-ph]}}.

\bibitem{Sirunyan:2018ouh}
{\bfseries CMS} Collaboration, A.~Sirunyan {\em et~al.}, ``{Measurements of
  Higgs boson properties in the diphoton decay channel in proton-proton
  collisions at $\sqrt{s} =$ 13 TeV},''
  \href{http://dx.doi.org/10.1007/JHEP11(2018)185}{{\em JHEP} {\bfseries 11}
  (2018) 185}, \href{http://arxiv.org/abs/1804.02716}{{\ttfamily
  arXiv:1804.02716 [hep-ex]}}.

\bibitem{Kazakov:2019wce}
D.~I. Kazakov, ``{RG Equations and High Energy Behaviour in Non-Renormalizable
  Theories},'' \href{http://dx.doi.org/10.1016/j.physletb.2019.134801}{{\em
  Phys. Lett. B} {\bfseries 797} (2019) 134801},
  \href{http://arxiv.org/abs/1904.08690}{{\ttfamily arXiv:1904.08690
  [hep-th]}}.

\bibitem{Kazakov:2020xbo}
D.~I. Kazakov, ``{Non-renormalizable Interactions: A Self-Consistency
  Manifesto},'' \href{http://arxiv.org/abs/2007.00948}{{\ttfamily
  arXiv:2007.00948 [hep-th]}}.

\bibitem{Hepp:1966eg}
K.~Hepp, ``{Proof of the Bogolyubov-Parasiuk theorem on renormalization},''
  \href{http://dx.doi.org/10.1007/BF01773358}{{\em Commun. Math. Phys.}
  {\bfseries 2} (1966) 301--326}.

\bibitem{Zimmermann:1969jj}
W.~Zimmermann, ``{Convergence of Bogolyubov's method of renormalization in
  momentum space},'' \href{http://dx.doi.org/10.1007/BF01645676}{{\em Commun.
  Math. Phys.} {\bfseries 15} (1969) 208--234}.

\bibitem{Abada:2007ux}
A.~Abada, C.~Biggio, F.~Bonnet, M.~B. Gavela, and T.~Hambye, ``{Low energy
  effects of neutrino masses},''
  \href{http://dx.doi.org/10.1088/1126-6708/2007/12/061}{{\em JHEP} {\bfseries
  12} (2007) 061},
\href{http://arxiv.org/abs/0707.4058}{{\ttfamily arXiv:0707.4058 [hep-ph]}}.

\bibitem{Aghanim:2018eyx}
{\bfseries Planck} Collaboration, N.~Aghanim {\em et~al.}, ``{Planck 2018
  results. VI. Cosmological parameters},''
  \href{http://arxiv.org/abs/1807.06209}{{\ttfamily arXiv:1807.06209
  [astro-ph.CO]}}.

\bibitem{Boyarsky:2014jta}
A.~Boyarsky, O.~Ruchayskiy, D.~Iakubovskyi, and J.~Franse, ``{Unidentified Line
  in X-Ray Spectra of the Andromeda Galaxy and Perseus Galaxy Cluster},''
  \href{http://dx.doi.org/10.1103/PhysRevLett.113.251301}{{\em Phys. Rev.
  Lett.} {\bfseries 113} (2014) 251301},
  \href{http://arxiv.org/abs/1402.4119}{{\ttfamily arXiv:1402.4119
  [astro-ph.CO]}}.

\bibitem{Deshpande:1977rw}
N.~G. Deshpande and E.~Ma, ``{Pattern of Symmetry Breaking with Two Higgs
  Doublets},''
\href{http://dx.doi.org/10.1103/PhysRevD.18.2574}{{\em Phys. Rev.} {\bfseries
  D18} (1978) 2574}.

\bibitem{Lee:1977eg}
B.~W. Lee, C.~Quigg, and H.~B. Thacker, ``{Weak Interactions at Very
  High-Energies: The Role of the Higgs Boson Mass},''
\href{http://dx.doi.org/10.1103/PhysRevD.16.1519}{{\em Phys. Rev.} {\bfseries
  D16} (1977) 1519}.

\bibitem{Gariazzo:2015rra}
S.~Gariazzo, C.~Giunti, M.~Laveder, Y.~Li, and E.~Zavanin, ``{Light sterile
  neutrinos},'' \href{http://dx.doi.org/10.1088/0954-3899/43/3/033001}{{\em J.
  Phys. G} {\bfseries 43} (2016) 033001},
  \href{http://arxiv.org/abs/1507.08204}{{\ttfamily arXiv:1507.08204
  [hep-ph]}}.

\bibitem{Hagstotz:2020ukm}
S.~Hagstotz, P.~F. de~Salas, S.~Gariazzo, M.~Gerbino, M.~Lattanzi, S.~Vagnozzi,
  K.~Freese, and S.~Pastor, ``{Bounds on light sterile neutrino mass and mixing
  from cosmology and laboratory searches},''
  \href{http://arxiv.org/abs/2003.02289}{{\ttfamily arXiv:2003.02289
  [astro-ph.CO]}}.

\bibitem{Casas:2001sr}
J.~Casas and A.~Ibarra, ``{Oscillating neutrinos and $\mu \to e, \gamma$},''
  \href{http://dx.doi.org/10.1016/S0550-3213(01)00475-8}{{\em Nucl. Phys. B}
  {\bfseries 618} (2001) 171--204},
  \href{http://arxiv.org/abs/hep-ph/0103065}{{\ttfamily arXiv:hep-ph/0103065}}.

\bibitem{Ibarra:2003up}
A.~Ibarra and G.~G. Ross, ``{Neutrino phenomenology: The Case of two
  right-handed neutrinos},''
  \href{http://dx.doi.org/10.1016/j.physletb.2004.04.037}{{\em Phys. Lett. B}
  {\bfseries 591} (2004) 285--296},
  \href{http://arxiv.org/abs/hep-ph/0312138}{{\ttfamily arXiv:hep-ph/0312138}}.

\bibitem{Bene__2005}
P.~Beneš, A.~Faessler, S.~Kovalenko, and F.~Šimkovic, ``Sterile neutrinos in
  neutrinoless double beta decay,''
  \href{http://dx.doi.org/10.1103/physrevd.71.077901}{{\em Physical Review D}
  {\bfseries 71} no.~7, (Apr, 2005) }.
  \url{http://dx.doi.org/10.1103/PhysRevD.71.077901}.

\bibitem{Obara:2017ndb}
{\bfseries KamLAND-Zen} Collaboration, S.~Obara, ``{Status of balloon
  production for KamLAND-Zen 800 kg phase},''
\href{http://dx.doi.org/10.1016/j.nima.2016.06.059}{{\em Nucl. Instrum. Meth.}
  {\bfseries A845} (2017) 410--413}.

\bibitem{Artusa:2014lgv}
{\bfseries CUORE} Collaboration, D.~R. Artusa {\em et~al.}, ``{Searching for
  neutrinoless double-beta decay of $^{130}$Te with CUORE},''
  \href{http://dx.doi.org/10.1155/2015/879871}{{\em Adv. High Energy Phys.}
  {\bfseries 2015} (2015) 879871},
\href{http://arxiv.org/abs/1402.6072}{{\ttfamily arXiv:1402.6072
  [physics.ins-det]}}.

\bibitem{Hartnell:2012qd}
{\bfseries SNO+} Collaboration, J.~Hartnell, ``{Neutrinoless Double Beta Decay
  with SNO+},'' \href{http://dx.doi.org/10.1088/1742-6596/375/1/042015}{{\em J.
  Phys. Conf. Ser.} {\bfseries 375} (2012) 042015},
\href{http://arxiv.org/abs/1201.6169}{{\ttfamily arXiv:1201.6169
  [physics.ins-det]}}.

\bibitem{Gomez-Cadenas:2013lta}
{\bfseries NEXT} Collaboration, J.~J. Gomez-Cadenas {\em et~al.}, ``{Present
  status and future perspectives of the NEXT experiment},''
  \href{http://dx.doi.org/10.1155/2014/907067}{{\em Adv. High Energy Phys.}
  {\bfseries 2014} (2014) 907067},
\href{http://arxiv.org/abs/1307.3914}{{\ttfamily arXiv:1307.3914
  [physics.ins-det]}}.

\bibitem{Barabash:2011aa}
A.~S. Barabash, ``{SeperNEMO double beta decay experiment},''
  \href{http://dx.doi.org/10.1088/1742-6596/375/1/042012}{{\em J. Phys. Conf.
  Ser.} {\bfseries 375} (2012) 042012},
\href{http://arxiv.org/abs/1112.1784}{{\ttfamily arXiv:1112.1784 [nucl-ex]}}.

\bibitem{KamLAND-Zen:2016pfg}
{\bfseries KamLAND-Zen} Collaboration, A.~Gando {\em et~al.}, ``{Search for
  Majorana Neutrinos near the Inverted Mass Hierarchy Region with
  KamLAND-Zen},'' \href{http://dx.doi.org/10.1103/PhysRevLett.117.109903,
  10.1103/PhysRevLett.117.082503}{{\em Phys. Rev. Lett.} {\bfseries 117} no.~8,
  (2016) 082503}, \href{http://arxiv.org/abs/1605.02889}{{\ttfamily
  arXiv:1605.02889 [hep-ex]}}.
[Addendum: Phys. Rev. Lett.117,no.10,109903(2016)].

\bibitem{Bhang:2012gn}
H.~Bhang {\em et~al.}, ``{AMoRE experiment: a search for neutrinoless double
  beta decay of Mo-100 isotope with Ca-40 MoO-100(4) cryogenic scintillation
  detector},''
\href{http://dx.doi.org/10.1088/1742-6596/375/1/042023}{{\em J. Phys. Conf.
  Ser.} {\bfseries 375} (2012) 042023}.

\bibitem{Tosi:2014zza}
{\bfseries EXO-200} Collaboration, D.~Tosi,
  \href{http://dx.doi.org/10.1142/9789814603164_0047}{``{The search for
  neutrino-less double-beta decay: summary of current experiments},''} in {\em
  {Proceedings, 14th ICATPP Conference on Astroparticle, Particle, Space
  Physics and Detectors for Physics Applications (ICATPP 2013): Como, Italy,
  September 23-27, 2013}}, pp.~304--314.
\newblock 2014.
\newblock
\href{http://arxiv.org/abs/1402.1170}{{\ttfamily arXiv:1402.1170 [nucl-ex]}}.
\newblock

\bibitem{Licciardi:2017oqg}
{\bfseries nEXO} Collaboration, C.~Licciardi, ``{The Sensitivity of the nEXO
  Experiment to Majorana Neutrinos},''
\href{http://dx.doi.org/10.1088/1742-6596/888/1/012237}{{\em J. Phys. Conf.
  Ser.} {\bfseries 888} no.~1, (2017) 012237}.

\bibitem{Alloul:2013bka}
A.~Alloul, N.~D. Christensen, C.~Degrande, C.~Duhr, and B.~Fuks, ``{FeynRules
  2.0 - A complete toolbox for tree-level phenomenology},''
  \href{http://dx.doi.org/10.1016/j.cpc.2014.04.012}{{\em Comput. Phys.
  Commun.} {\bfseries 185} (2014) 2250--2300},
\href{http://arxiv.org/abs/1310.1921}{{\ttfamily arXiv:1310.1921 [hep-ph]}}.

\bibitem{Belanger:2018mqt}
G.~Bélanger, F.~Boudjema, A.~Goudelis, A.~Pukhov, and B.~Zaldivar,
  ``{micrOMEGAs5.0 : Freeze-in},''
  \href{http://dx.doi.org/10.1016/j.cpc.2018.04.027}{{\em Comput. Phys.
  Commun.} {\bfseries 231} (2018) 173--186},
\href{http://arxiv.org/abs/1801.03509}{{\ttfamily arXiv:1801.03509 [hep-ph]}}.

\bibitem{Aprile:2012nq}
{\bfseries XENON100} Collaboration, E.~Aprile {\em et~al.}, ``{Dark Matter
  Results from 225 Live Days of XENON100 Data},''
  \href{http://dx.doi.org/10.1103/PhysRevLett.109.181301}{{\em Phys. Rev.
  Lett.} {\bfseries 109} (2012) 181301},
  \href{http://arxiv.org/abs/1207.5988}{{\ttfamily arXiv:1207.5988
  [astro-ph.CO]}}.

\bibitem{Aprile:2016swn}
{\bfseries XENON100} Collaboration, E.~Aprile {\em et~al.}, ``{XENON100 Dark
  Matter Results from a Combination of 477 Live Days},''
  \href{http://dx.doi.org/10.1103/PhysRevD.94.122001}{{\em Phys. Rev. D}
  {\bfseries 94} no.~12, (2016) 122001},
  \href{http://arxiv.org/abs/1609.06154}{{\ttfamily arXiv:1609.06154
  [astro-ph.CO]}}.

\bibitem{Akerib:2019diq}
{\bfseries LUX} Collaboration, D.~Akerib {\em et~al.}, ``{First direct
  detection constraint on mirror dark matter kinetic mixing using LUX 2013
  data},'' \href{http://dx.doi.org/10.1103/PhysRevD.101.012003}{{\em Phys. Rev.
  D} {\bfseries 101} no.~1, (2020) 012003},
  \href{http://arxiv.org/abs/1908.03479}{{\ttfamily arXiv:1908.03479
  [hep-ex]}}.

\bibitem{Bernabei:2010mq}
{\bfseries DAMA, LIBRA} Collaboration, R.~Bernabei {\em et~al.}, ``{New results
  from DAMA/LIBRA},''
  \href{http://dx.doi.org/10.1140/epjc/s10052-010-1303-9}{{\em Eur. Phys. J. C}
  {\bfseries 67} (2010) 39--49},
  \href{http://arxiv.org/abs/1002.1028}{{\ttfamily arXiv:1002.1028
  [astro-ph.GA]}}.

\bibitem{Aalseth:2012if}
{\bfseries CoGeNT} Collaboration, C.~Aalseth {\em et~al.}, ``{CoGeNT: A Search
  for Low-Mass Dark Matter using p-type Point Contact Germanium Detectors},''
  \href{http://dx.doi.org/10.1103/PhysRevD.88.012002}{{\em Phys. Rev. D}
  {\bfseries 88} (2013) 012002},
  \href{http://arxiv.org/abs/1208.5737}{{\ttfamily arXiv:1208.5737
  [astro-ph.CO]}}.

\bibitem{Agnese:2013rvf}
{\bfseries CDMS} Collaboration, R.~Agnese {\em et~al.}, ``{Silicon Detector
  Dark Matter Results from the Final Exposure of CDMS II},''
  \href{http://dx.doi.org/10.1103/PhysRevLett.111.251301}{{\em Phys. Rev.
  Lett.} {\bfseries 111} no.~25, (2013) 251301},
  \href{http://arxiv.org/abs/1304.4279}{{\ttfamily arXiv:1304.4279 [hep-ex]}}.

\bibitem{Hooper:2010mq}
D.~Hooper and L.~Goodenough, ``{Dark Matter Annihilation in The Galactic Center
  As Seen by the Fermi Gamma Ray Space Telescope},''
  \href{http://dx.doi.org/10.1016/j.physletb.2011.02.029}{{\em Phys. Lett. B}
  {\bfseries 697} (2011) 412--428},
  \href{http://arxiv.org/abs/1010.2752}{{\ttfamily arXiv:1010.2752 [hep-ph]}}.

\bibitem{Aguilar:2013qda}
{\bfseries AMS} Collaboration, M.~Aguilar {\em et~al.}, ``{First Result from
  the Alpha Magnetic Spectrometer on the International Space Station: Precision
  Measurement of the Positron Fraction in Primary Cosmic Rays of
  0.5\textendash{}350 GeV},''
  \href{http://dx.doi.org/10.1103/PhysRevLett.110.141102}{{\em Phys. Rev.
  Lett.} {\bfseries 110} (2013) 141102}.

\bibitem{Cholis:2008qq}
I.~Cholis, D.~P. Finkbeiner, L.~Goodenough, and N.~Weiner, ``{The PAMELA
  Positron Excess from Annihilations into a Light Boson},''
  \href{http://dx.doi.org/10.1088/1475-7516/2009/12/007}{{\em JCAP} {\bfseries
  12} (2009) 007}, \href{http://arxiv.org/abs/0810.5344}{{\ttfamily
  arXiv:0810.5344 [astro-ph]}}.

\bibitem{Parker_2018}
R.~H. Parker, C.~Yu, W.~Zhong, B.~Estey, and H.~Müller, ``Measurement of the
  fine-structure constant as a test of the standard model,''
  \href{http://dx.doi.org/10.1126/science.aap7706}{{\em Science} {\bfseries
  360} no.~6385, (Apr, 2018) 191–195}.
  \url{http://dx.doi.org/10.1126/science.aap7706}.

\bibitem{Granelli:2020pim}
A.~Granelli, K.~Moffat, Y.~Perez-Gonzalez, H.~Schulz, and J.~Turner,
  ``{ULYSSES: Universal LeptogeneSiS Equation Solver},''
  \href{http://arxiv.org/abs/2007.09150}{{\ttfamily arXiv:2007.09150
  [hep-ph]}}.

\bibitem{Blanchet:2011xq}
S.~Blanchet, P.~Di~Bari, D.~A. Jones, and L.~Marzola, ``{Leptogenesis with
  heavy neutrino flavours: from density matrix to Boltzmann equations},''
  \href{http://dx.doi.org/10.1088/1475-7516/2013/01/041}{{\em JCAP} {\bfseries
  01} (2013) 041}, \href{http://arxiv.org/abs/1112.4528}{{\ttfamily
  arXiv:1112.4528 [hep-ph]}}.

\bibitem{Bechtle:2013xfa}
P.~Bechtle, S.~Heinemeyer, O.~St\r{a}l, T.~Stefaniak, and G.~Weiglein,
  ``{$HiggsSignals$: Confronting arbitrary Higgs sectors with measurements at
  the Tevatron and the LHC},''
  \href{http://dx.doi.org/10.1140/epjc/s10052-013-2711-4}{{\em Eur. Phys. J. C}
  {\bfseries 74} no.~2, (2014) 2711},
  \href{http://arxiv.org/abs/1305.1933}{{\ttfamily arXiv:1305.1933 [hep-ph]}}.

\end{thebibliography}\endgroup

\end{document}